\documentclass[fleqn,usenatbib]{mnras}

\usepackage{newtxtext,newtxmath,gensymb,pdflscape}
\usepackage{float}

\usepackage[T1]{fontenc}

\DeclareRobustCommand{\VAN}[3]{#2}
\let\VANthebibliography\thebibliography
\def\thebibliography{\DeclareRobustCommand{\VAN}[3]{##3}\VANthebibliography}

\usepackage{graphicx}	
\usepackage{amsmath}	

\newcommand{\psra}{PSR~J0101$-$6422}
\newcommand{\psrb}{PSR~J1101$-$6424}
\newcommand{\psrc}{PSR~J1514$-$4946}
\newcommand{\psrd}{PSR~J1614$-$2230}
\newcommand{\psre}{PSR~J1732$-$5049}
\newcommand{\psrf}{PSR~J1909$-$3744}
\newcommand{\psrg}{PSR~J1125$-$6014}

\title[Timing analysis of seven pulsars with MeerKAT]{Searches for Shapiro delay in seven binary pulsars using the MeerKAT telescope}

\author[M. Shamohammadi et al.]{
M.~Shamohammadi,$^{1,2}$\thanks{E-mail: \href{mailto:msh.ph.ir@gmail.com}{msh.ph.ir@gmail.com}}
M.~Bailes,$^{1,2}$
P.~C.~C.~Freire,$^{3}$
A.~Parthasarathy,$^{3}$
D.~J.~Reardon,$^{1,2}$
R.~M.~Shannon,$^{1,2}$\newauthor
V.~Venkatraman~Krishnan,$^{3}$
M.~C.~i.~Bernadich,$^{3}$
A.~D.~Cameron,$^{1,2}$
D.~J.~Champion,$^{3}$
A.~Corongiu,$^{5}$\newauthor
C.~Flynn,$^{1,2}$
M.~Geyer,$^{6}$
M.~Kramer,$^{3}$
M.~T.~Miles,$^{1,2}$
A.~Possenti,$^{5}$
R.~Spiewak$^{1,2,4}$
\\
$^{1}$Centre for Astrophysics and Supercomputing, Swinburne University of Technology, PO Box 218, Hawthorn, VIC 3122, Australia \\
$^{2}$ARC Centre of Excellence for Gravitational Wave Discovery (OzGrav), Mail H29, Swinburne University of Technology, PO Box 218, Hawthorn, VIC 3122, Australia \\
$^{3}$Max-Planck-Institut f\"{u}r Radioastronomie, Auf dem H\"{u}gel 69, D-53121 Bonn, Germany \\
$^{4}$Jodrell Bank Centre for Astrophysics, Department of Physics and Astronomy, University of Manchester, Manchester M13 9PL, UK\\
$^5$INAF--Osservatorio Astronomico di Cagliari, Via della Scienza 5, I-09047 Selargius (CA), Italy\\
$^{6}$South African Radio Astronomy Observatory, 2 Fir Street, Black River Park, Observatory 7925, South Africa\\
}

\date{Accepted XXX. Received YYY; in original form ZZZ}

\pubyear{2022}

\begin{document}
\label{firstpage}
\pagerange{\pageref{firstpage}--\pageref{lastpage}}
\maketitle

\begin{abstract}


Precision timing of millisecond pulsars in binary systems enables
observers to detect the relativistic Shapiro delay induced by space time curvature. When favourably aligned, this enables constraints to be placed on the component masses and system orientation. 
Here we present the results of timing campaigns on seven binary millisecond pulsars observed with the 64-antenna MeerKAT radio telescope that show evidence of Shapiro delay: PSRs~J0101$-$6422, J1101$-$6424, J1125$-$6014, J1514$-$4946, J1614$-$2230, J1732$-$5049, and J1909$-$3744. Evidence for Shapiro delay was found in all of the systems, and for three the orientations and data quality enabled strong constraints on their orbital inclinations and component masses. 
For PSRs~J1125$-$6014, J1614$-$2230 and J1909$-$3744, we determined pulsar masses to be $M_{\rm p} = 1.68\pm 0.17 \, {\rm M_{\odot}} $, $1.94\pm 0.03 \, {\rm M_{\odot}} $ and $1.45 \pm 0.03 \, {\rm M_{\odot}}$, and companion masses to be $M_{\rm c} = 0.33\pm 0.02 \, {\rm M_{\odot}} $, $0.495\pm 0.005 \, {\rm M_{\odot}} $ and $0.205 \pm 0.003 \, {\rm M_{\odot}}$, respectively. This provides the first independent confirmation of PSR~J1614$-$2230's mass, one of the highest known. 
The Shapiro delays measured for PSRs~J0101$-$6422, J1101$-$6424, J1514$-$4946, and J1732$-$5049 were only weak, and could not provide interesting component mass limits. Despite a large number of millisecond pulsars being routinely timed, relatively few have accurate masses via Shapiro delays. We use simulations to show that this is expected, and provide a formula for observers to assess how accurately a pulsar mass can be determined. We also discuss the observed correlation between pulsar companion masses and spin period, and the anti-correlation between recycled pulsar mass and their companion masses.

\end{abstract}




\section{INTRODUCTION}

Radio pulsars are rapidly rotating neutron stars, with radio emission arriving at our observatories as a highly predictable train of pulses. The group of pulsars that have a spin period of $P \leq 30$ ms and a rate of spin-down of $\dot{P} \leq 10^{-17}$ are often called millisecond pulsars (MSPs). This subclass is thought to be produced through mass transfer from an evolving giant companion star (during the Roche-lobe overflow (RLO) stage) onto the neutron star. The consequent transfer of angular momentum during this stage decreases the spin period of the pulsar, reduces the magnetic field strength and circularises its orbit \citep{rs82,acr+82,l08,th23}.

Owing to their high rotational stability, MSPs have been used as astrophysical tools in Pulsar Timing Arrays (PTAs) to search for ultra-low (nanohertz) frequency gravitational waves \citep{srll+15,hd17,abb+20,gsr+21,cbp+22,aab+22}. In some binary pulsars, we can detect relativistic effects in the orbital motion and the propagation of light in the spacetime of the system, these are quantified by the so-called `post-Keplerian' (PK) parameters.

With two PK parameters, we can use a theory of gravity to derive the masses of the pulsar and the companion object, as well as the orbital inclination angle with respect to our line of sight \citep{kkdt13,of16}. Measuring pulsar masses helps us to determine the underlying distribution of neutron star masses (see also \citealt{ato+16}). The neutron stars with highest masses probe the Equation of State (EOS) of nuclear matter \citep{lp04} (for instance PSR~J1614$-$2230 \citealt{abb+18}, PSR~J0348$+$0432 \citealt{afw+13} and PSR~J0740$+$6620 \citealt{fcp+21}), since any EOS that fails to predict such massive neutron stars can be ruled out \citep{lp04,of16}\footnote{We note that even more massive neutron stars could exist (e.g., \citealt{rkf+21,rkf+22}), but such masses were not obtained from the measurement of PK parameters, and rely on assumptions about surface heating and reradiation and are more model-dependent as a result.}.

If we measure more than two PK parameters, we can test the theory of gravity being used to 
interpret them, by verifying whether each pair of PK parameters yields consistent mass estimates \citep{s03,ksm+06,k14,v19,ksm+21,f22}. In some extreme cases, we can even look for other relativistic effects like spin-orbit coupling \citep{wrt89,sta04,fst14,kbs20}, which might, in some cases, provide additional tests of GR and even constrain the neutron star EOS \citep{hkw+20}.

In pulsar timing, it is a standard practice to develop a timing model via an iterative procedure that can accurately predict the time of arrival of every pulse (ToA). The timing model includes parameters that describe the pulsar's spin, astrometric, and orbital properties; however, there are also noise processes that can be correlated with parameters in the timing model. The timing model is then iteratively updated using software such as \textsc{tempo2} \citep{hem06,ehm06}, which is used to calculate and analyse timing residuals. To properly analyse the arrival times, we need to account for the noise processes, which for the purposes of this analysis, are limited to the systematic white noise, achromatic red noise, and that induced by dispersion measure variations. 

According to the general theory of relativity, the Shapiro delay is the additional delay in the propagation of any signal travelling at the speed of light caused by the space-time curvature near massive objects \citep{s64}. If a pulsar's orbit is observed close to edge-on, the Shapiro delay caused by the space-time curvature near the companion can be detected in the pulsar signal; in such cases it can be used to constrain the component masses and system's orbital inclination. The shape of the timing residuals as a function of orbital phase determines the orientation, and the amplitude of the delay peak is proportional to the companion mass.

In this paper, we present timing and noise analyses of seven binary MSPs: PSR~J0101$-$6422, J1101$-$6424, J1125$-$6014, J1514$-$4946, J1614$-$2230, J1732$-$5049, and J1909$-$3744, using observations from the 64-antenna MeerKAT radio telescope.

Among these MSPs, PSRs~J1909$-$3744 and J1614$-$2230 are of great importance.
PSR~J1909$-$3744 has an extremely narrow pulse \citep{jbk+03}. For that reason, it has the highest timing precision of any pulsars known \citep{msb+22}. This, in combination with the high orbital inclination resulted in the first precise mass measurement for an MSP \citep{jhb+05}. The high timing precision means
it has been timed by all the major timing arrays \citep{src+13,ltm+15,abb+20}; this has resulted in multiple independent measurements of its mass \citep{abb+18,lgi+20,rsc+21}.

For more than a decade, PSR~J1614$-$2230 has been one of the most important pulsars for constraining EOSs due to its high mass. \cite{dpr+10} found the mass of the pulsar to be $1.97 \pm 0.04 \, {\rm M_{\odot}}$ by measuring the amplitude of the Shapiro time delay with the Green Bank Telescope. This mass has been refined since \citep{fpe+16,abb+18}, but these masses have never been confirmed with data from other radio telescopes.

The MeerTime Pulsar Timing Array (MPTA) has been established to frequently observe up to 89 pulsars with extremely high precision \citep{msb+22} with the aims of searching for gravitational waves coming from merging supermassive black hole binaries, testing theories of gravity, studying the structure of interstellar plasma, exploring the interior structure of neutron stars, and understanding the origin and evolution of MSP binary systems \footnote{\url{www.meertime.org}}. 
By augmenting the MPTA data set with
intense observing campaigns around
superior conjunction, the ongoing Relativistic Binary programme \citep{ksv21} under the MeerTime Large Survey Project with the MeerKAT telescope \citep{bja+20} is aiming to increase the number of known pulsar masses amongst other objectives (e.g. \citet{svf+22}). This paper, as a part of the Relativistic Binary programme, investigates measurements of Shapiro delay and (where possible) the derived masses of seven MSPs and their companion stars.
In Section \ref{sec:data_set}, we describe the MeerKAT observations and the data processing leading to the calculation of the ToAs. 
In Section \ref{sec:methods}, we present our methods for modelling noise processes, updating the timing models, and measuring the Shapiro delays.
In Section \ref{sec:results}, we present our results on constrained masses and orbital inclination angles. We also present two simulations performed for exploring the change in the pulsar mass error as a function of the orbital inclination angle. 
In Section \ref{sec:discussion}, we discuss our measurements and compare them to previous values, finding good agreement and
explore correlations between the component masses and the spin periods of recycled pulsars.
In Section \ref{sec:conclusions}, we summarise the major results and draw our conclusions.

\section{THE DATA SET}\label{sec:data_set}

The observations of the MSP sample began in early 2019 and concluded in early 2022. Observations were also conducted as part of the MeerTime Pulsar Timing Array project (Miles et al. 2022, in press). The observations were performed with the 64$-$dish MeerKAT radio telescope using two receivers: the Ultra-High Frequency (UHF) receiver, operating between 544-1088 $\rm MHz$, and the L-band receiver, operating between 856-1712 $\rm MHz$, both with 1024 frequency channels \citep{bja+20}. Using the two receivers, we collected coherently de-dispersed data folded in real time at the apparent pulsar period; at each observing session only one receiver was used. PSRs~J1101$-$6424, J1125$-$6014, J1514$-$4946, J1614$-$2230, and J1909$-$3744 have been observed with the L-band receiver only. 
UHF observations of PSR~J0101$-$6422 and J1732$-$5049 were included as these had relatively steep 
spectral indices of $-$1.7 and $-$1.9, respectively, (Gitika et al. in prep) and benefitted from UHF observing
campaigns near superior conjunction.
The minimum number of 37 observations was for PSR~J1514$-$4946 and the maximum number of 187 observations was for J1909$-$3744, over a three year time span, and have good orbital coverage. For each pulsar, we conducted a long, targeted observing campaign, centred in time near superior conjunction (where the companion passes between the Earth and the pulsar), to optimise our sensitivity to the Shapiro delay. A summary of the data spans, total integration times, and the number of observations for each pulsar are listed in Tables \ref{tab:timing_solution_1} and \ref{tab:timing_solution_2}.

Frequency channels affected with radio-frequency interference (RFI) were zero-weighted with the \textsc{rfihunter} RFI rejection tool\footnote{\url{https://github.com/mshamohammadi/rfihunter}}. This software subtracts a normalised and phase-aligned mean pulse profile from the data and creates pulse profile residuals. To do this, it fits a smoothed integrated pulse profile to each 8-sec integrated frequency channel, using a non-linear least squares fitting algorithm. The pulse residuals are then searched for the detection of RFI-affected frequency channels, after normalising with a `median bandpass model'. Following this, the frequency channels with statistics above $3.5\sigma$ from the median are zero-weighted, similar to the method that was employed in \cite{bja+20}. The statistics used for RFI excision are 1) the maximum amplitude of the Fourier transform of the mean-subtracted residuals, 2) the standard deviation, and 3) the range (following the statistics used in \cite{lkg+16}). To construct a median bandpass model of each receiver, we used the off-pulse baselines and used the Gaussian Process Regression module\footnote{\url{https://scikit-learn.org/stable/modules/gaussian_process.html}} implemented in \textsc{scikit-learn} package \citep{scikit-learn} to determine a robust median.

After RFI excision, we installed pulsar ephemerides updated for the work of \citet[][private communication]{sbm+22}, averaged the observations in time and reduced the data to 16 frequency channels. This channelisation allowed us to take into account the temporal variations of pulsar dispersion-measure (DM) \citep{kcsh+13,jml+17}, as well as the detection of the in-band profile evolution, which requires frequency-dependent parameters in our timing solution \citep{dfgn+13}.

For every pulsar, we summed high signal-to-noise observations in time, polarization, and frequency, and then using wavelet smoothing techniques implemented in the \textsc{psrchive} program \texttt{psrsmooth}, we created a noise-free pulse profile, often termed a `standard' or `template'. For PSRs~J1614$-$2230 and J1909$-$3744, we used PulsePortraiture\footnote{\url{https://github.com/pennucci/PulsePortraiture/tree/py3}} to create a frequency-dependent template, which is a profile that evolves in frequency, referred to as a `portrait' \citep{p19}, the aim of which is to account for frequency-dependent time delays caused by profile evolution. This was possible
because we had very high signal-to-noise ratio observations of these two pulsars,
making the profile evolution with frequency secure.

Finally, we used the Fourier domain Monte Carlo (FDM) algorithm in \texttt{pat}, which fits for a phase gradient between the Fourier transforms of the template and profile, to calculate a ToA for every pulse profile, and selected the profiles with a signal-to-noise ratio higher than $10$. This algorithm uses a Markov Chain Monte Carlo method to determine the likely ToA uncertainties. \footnote{The algorithm is in the
ProfileShiftFit.C source code from the psrchive sourceforge.net repository.}

\section{METHODS}\label{sec:methods}
For standard pulsar timing practices, we need a timing model (solution) that includes pulsar parameters such as its astrometry, spin, dispersion measure, binary motion (five Keplerian parameters), and a Solar System model. The astrometric parameters include position (Right Ascension $\alpha$ and Declination $\delta$ ), proper motion ($\mu_{\alpha}\cos \delta$, $\mu_{\delta}$), and when significant, a parallax. When profiles possess strong profile evolution as a function of frequency, radio frequency derivative parameters are added to the timing model. A list of systematic jumps are also added to the timing model if needed between frequency bands or observational set-ups. We commenced our timing with the latest pulsar ephemerides from \citet[][private communication]{sbm+22} as the initial timing models, and added two Shapiro parameters (will be discussed below) into the models, and refined these based upon our ToAs using \textsc{tempo2}, utilizing a generalized least-squares fitting method to minimize the timing residuals, and updated them constantly by adding ToAs from new observations. We used JPL DE440 as our Solar System ephemeris \citep{pfw+21}, the Bureau International des Poids et Mesures BIPM2020 as our realization of terrestrial time, and Barycentric Coordinate Time (TCB) for our units.

\subsection{Timing model}

For a pulsar in a binary with a low eccentricity, $e$, it is difficult to measure the longitude of periastron, $\omega$, and there is a high correlation between the longitude of periastron and the time of periastron, $T_{0}$. In such systems, there is typically a high correlation between the orbital period, $P_{\rm b}$, and the time-derivative of the longitude of periastron, $\dot{\omega}$. The ELL1 binary model can address this problem in systems with a low $e$ \citep{lcw01}. Instead of $e$, $\omega$, and $T_{0}$, this model uses the time of passage through the ascending node, $T_{\rm asc}$ , and the two Laplace-Lagrange parameters, $\eta \equiv e \sin \omega$ and $\kappa \equiv e \cos \omega$, which can be measured with less covariance.  

The effects of Shapiro delay have been parameterised in two binary models implemented in \textsc{tempo2}: DD \citep{dd86} and ELL1 using a function of two post-Keplerian (PK) parameters: range ($r$) and shape ($s$) in the following forms:
\begin{align}
     & r = T_{\odot} \, M_{\rm c} \\
     & s = \sin i
\end{align}
where $M_{\rm c}$ is the mass of the companion and $T_{\odot} \equiv G \, {\rm M_{\odot}} \, c^{-3} = 4.925490947 \mu$s is the mass of the Sun in time units.
The Shapiro delay, $\Delta_{\rm s}$, in the ELL1 binary model is modelled as
\begin{equation}\label{eq:shapiro_formula_classic}
    \Delta_{\rm s} = -2r \ln{(1 - s \sin \Phi)},
\end{equation}
where $\Phi$ is the celestial longitude of the binary. At high inclination angles, where $\sin i\sim 1$, the peak of the Shapiro delay is very pronounced, but at other inclinations it becomes extremely covariant with the R{\"o}mer delay (caused by the pulsar's orbital motion, \citealt{lcw01}) from the circular orbit.

The problem with $r$ and $s$ is that they become highly correlated, in particular at low (i.e. face-on) orbital inclinations.
Because of this, \cite{fw10} parameterised the Shapiro delay for low-eccentric binary pulsars in Fourier space, using the sum of the harmonics of the Shapiro delay. They introduced the two new PK parameters that are proportional to the third and fourth harmonic amplitudes ($h_3$,$h_4$). For higher inclinations, the best parameterisation is $h_3$ and $\varsigma = h_{4} / h_{3}$, because the latter parameter can be measured more precisely than $h_4$. The main advantage of these ``orthometric" parameters is that they are less correlated with one another compared to $r$ and $s$, thus providing an optimal description of the $m_{\rm c}, \sin i$ constraints provided by the Shapiro delay. The orthometric parameters relate to $r,s$ according to the following relations \citep{fw10}: 
\begin{align}
     & \varsigma = \frac{s}{1+  \sqrt{1-s^{2}} } \\
     & h_{\rm 3} = r \varsigma^{3},
\end{align}
and obtained the equivalent Shapiro delay formula 
\begin{equation}\label{eq:shapiro_formula_orthometric}
    \Delta_{\rm s} = -\frac{2h_{3}}{\varsigma^{3}} \ln{(1 + \varsigma^{2} - 2\varsigma \, \sin \Phi)}.
\end{equation}
Within the range of possible orbital inclinations ($0^\circ \leq i \leq 180^\circ$), we have $0 \leq \sin i \leq 1$; this implies that $0 \leq \varsigma \leq 1$ and $h_{3}>0$ (since $r \equiv T_{\odot} \, M_{\rm c}$ must be positive).

Three new binary models were established based upon the ($h_3$,$h_4$) and ($h_3$,$\varsigma$) parameterisations and implemented in the ELL1H binary model as an alternative to the ELL1 binary model.
In this work, we have used these models with the $h_3$ and $\varsigma$ parameterisation when searching for the relativistic Shapiro delay in the timing residuals for each pulsar because the binaries being studied have relatively high orbital inclinations.
We note that, although physical constraints require that $0 \leq \varsigma \leq 1$ and $h_3 > 0$, such constraints are not enforced by the ELL1H orbital model, in the same way that the conditions $0 \leq s \leq 1$ and $r > 0$ are not enforced by any orbital model when fitting for $r$ and $s$.

\subsection{Noise models}
There are two basic contributions in calculating timing residuals; the deterministic and stochastic noise components \citep{lah+14}. The latter can be modelled using three white and two red stochastic processes:
\begin{enumerate}
    \item White (time uncorrelated) noise: can be described using three components as follows:
    \begin{enumerate}
        \item EFAC: a factor $\alpha_{\rm i}$ that is multiplied to any ToA uncertainty $\sigma_{\rm i}$, where $i$ is the i-th ToA. One possible contribution to this term is due to the radiometer noise after measuring ToA uncertainty using the template matching algorithm. For every receiver, backend, or pulsar, we might need to use different EFAC parameter.
        \item EQUAD: an error term $\beta_{\rm i}$ that is added in quadrature to the scaled ToA uncertainty by EFAC. It is usually necessary to have separate EQUAD parameters for different receiver, backend, or pulsar. Therefore, the uncertainty of the i-th ToA scaled with EFAC and EQUAD can be written as
            \begin{equation}\label{eq:efac_equad}
            \begin{array}{cc}
                 & {\sigma}_{\mathrm{i, \, whitened}}^{2} =  \left(\alpha_{\mathrm{i}} \sigma_{\mathrm{i}}\right)^{2}+\beta_{i}^{2}. \\
                 \end{array}
            \end{equation}
        \item ECORR: a noise term that describes the correlations between the sub-bands of one observation, and is not correlated between observations taken at different epochs. We used the TECORR parameter described in Equation 1 in \cite{bja+20} in our noise analysis.
    \end{enumerate}

    \item Red noise: This term accounts an achromatic stationary Gaussian process with a low-frequency, and is modeled using a power-law spectrum that has a form of

        \begin{equation}
            P_{\rm red}(f)=A_{\rm red}^{2}\left ( \frac{f}{1 {\rm yr}^{-1}} \right )^{\gamma_{\rm red}}
        \end{equation}
    where $S_{\rm red}(f)$ is the red noise power spectral density at frequency $f$. $A_{\rm red}$ is the amplitude of the red noise in $\mu$s yr$^{1/2}$, and $\gamma_{\rm red}$ is the spectral index of the noise process.

    \item Dispersion noise:  pulses at higher frequencies arrive earlier in a dispersive interstellar medium (ISM). The delay due to the frequency dispersion is measured (in seconds) using  $1/2.41\times10^{-4} \, DM / \nu^{2}$, where DM is the dispersion measure and $\nu$ is the observing frequency in MHz. DM, measured in $\rm pc \, cm^{-3}$, is the integral of the electron number density along our line of sight 
    \begin{equation}
        DM = \int_{0}^{L} n_{\mathrm{e}}(z) dz,
    \end{equation}
    where $L$ is the distance to the pulsar.  
    Our line of sight changes due to the motion of the pulsar and the Earth, causing changes in the DM with time. Additionally, the turbulent ISM between the pulsar and the Earth has a bulk motion, which induces additional changes in the DM with time. Therefore, DM is epoch dependent, and DM corrections are required after every observation. Through the use of broad band observations, it is possible to mitigate the effects of dispersion measure variations.
    
    These temporal DM variation introduces an additional noise term in the timing residuals. We describe this temporal variation using a power-law spectrum that has a form similar to the red noise spectrum,
        \begin{equation}
            P_{\rm DM}(f)=A_{\rm DM}^{2}\left ( \frac{f}{1 yr^{-1}} \right )^{\gamma_{\rm DM}} \left ( \frac{\nu}{\nu_{\rm c}} \right )^{-2},
        \end{equation}
    where $A_{\rm DM}$ is the DM noise amplitude that scales with $\nu^{-2}$, and $\nu_{\rm c}$ is the centre frequency of the observation.

\end{enumerate}

\subsection{Noise analysis}
We perform our noise analysis with \textsc{temponest} \citep{lah+14}, utilising an efficient Bayesian inference sampler, Multinest \citep{fh08,fhb09}, for exploring the parameter space and producing marginal posterior distributions of parameters. Model comparison between two timing models can be quantified using the ratio of their evidences, which is called the Bayes factor $\mathcal{B}$. Using the evidences returned by \textsc{temponest}, we calculate the log Bayes factor, $\ln{\mathcal{B}}$, and define significant support for one model in comparison to another model as $\mathcal{B} \geq 10$ \citep{kr95}, or $\ln{\mathcal{B}} \geq 2.3$.

\subsection{Relativistic parameter derivations}
There are additional physical parameters that can be derived from the measured parameters in the pulsar ephemerides. We can constrain the pulsar mass, $M_{\rm p}$, the companion mass, $M_{\rm c}$, and the orbital inclination angle, $i$, by measuring the relativistic Shapiro delay parameters, $h_{3}$ and $\varsigma$, in our timing residuals. Every binary system has a precise binary mass function, $f_{\rm m}$:

\begin{equation}\label{eq:mass_func}
f_{\rm m}=\frac{\left(M_{\rm c} \, \sin i\right)^{3}}{\left(M_{\rm p}+M_{\rm c}\right)^{2}}=\frac{4 \pi^{2}}{G} \frac{\left(a_{\rm p} \, \sin i\right)^{3}}{P_{\rm b}^{2}}=\frac{4 \pi^{2}}{T_{\odot}} \frac{x^{3}}{P_{\rm b}^{2}},
\end{equation}
which relates the unknown masses and the orbital inclination angle of the system. The mass function can be calculated using the observed orbital period and the observed projected semi-major axis of the pulsar, $x=a_{\rm p} \sin i /c$ in units of light-seconds, where $a_{\rm p}$ is the semi-major axis of the pulsar, and $c$ is the speed of light. The quantity $x$ can also experience a secular change, and it can experience a periodic modulation due to the orbital motion of the Earth around the Sun. This annual motion of the Earth causes the observed orbital inclination angle to have an annual periodic change that can be obtained by measuring the annual orbital parallax. In the case of the apparent change in $x$ due to the proper motion, from \cite{k96}, the maximum value of $\dot{x}/x$ provides an upper limit of
\begin{equation}\label{eq:xdot}
i \leq \tan^{-1}\left [ 1.54 \times 10^{-16} \left ( \frac{\mu_{T}}{\rm mas ~ yr^{-1}} \right )  / \left ( \frac{\dot{x}}{x} \right )_{\rm max} \right ]
\end{equation}
for the orbital inclination angle, where $\mu_{\rm T}$ is the total proper motion of the system in $\rm mas ~ yr^{-1}$. This provides another way to constrain the system inclination angle using the $\dot{x}$ measurement. 

In order to constrain the pulsar mass through the measurement of the Shapiro delay, it is necessary to constrain both the companion mass and the system inclination angle. So, the error in the pulsar mass $\sigma_{\rm M_{p}}$, depends on the errors in both the companion mass $\sigma_{\rm M_{c}}$ and the sine of the inclination angle $\sigma_{\rm \sin i}$, and can be estimated using partial derivatives of the mass function (Eq. \ref{eq:mass_func}), 
\begin{equation}\label{eq:mp_error}
\sigma_{\rm M_{p}}\approx \sqrt{A^{2} \, \sigma_{\rm M_{c}}^{2} + B^{2} \, \sigma_{\rm \sin i}^{2} + 2\, AB\, \operatorname{Cov}[M_{\rm c},\sin i]}
\end{equation}
\\
where
\begin{equation}\label{eq:mp_error_1}
A = \frac{3 M_{\rm c}^{1 / 2} \, {\sin i}^{3 / 2}}{2f_{\rm m}^{1 / 2}} - 1,
\end{equation}
\begin{equation}\label{eq:mp_error_2}
B = \frac{3 M_{\rm c}^{3 / 2} \, {\sin i}^{1 / 2}}{2f_{\rm m}^{1 / 2}}, 
\end{equation} 
and $\operatorname{Cov}[M_{\rm c},\sin i]$ is the covariance between $M_{\rm c}$ and $\sin i$.
We note that this equation was used for an efficient estimate of the uncertainties of $M_p$ in the simulations, not to estimate the uncertainties of $M_p$ for the pulsars in his work.

\section{RESULTS}\label{sec:results}
 
A summary of spin, astrometric, orbital, and noise parameters for each pulsar is listed in the Tables \ref{tab:timing_solution_1} and \ref{tab:timing_solution_2} in Appendix \ref{sec:ephemerides}. For PSRs~J1125$-$6014, J1614$-$2230 and J1909$-$3744, we report measurements of orthometric Shapiro parameters of $h_{3}$ and $\varsigma$, and from these we have derived pulsar masses, the companion masses, and the orbital inclination angles. Figures \ref{fig:J1125_PDF}, \ref{fig:J1614_PDF}, and \ref{fig:J1909_PDF} provide the two-dimensional posteriors of the measured and the derived parameters. Among the seven pulsars, we have significantly measured the pulsar and the companion star masses for PSR~J1125$-$6014, J1614$-$2230, and J1909$-$3744. These are described in section \ref{sec:mass_measurement} and summarised in the Table \ref{tab:mass_inclination}. Considering the posterior probability distributions of the derived parameters, the values and their uncertainties are taken from the cumulative distribution function at the values of $0.16$ (lower bound), $0.5$ (median), and $0.84$ (upper bound). We define the orbital inclination angles of the MSP binary systems to be in the range of $0^{\circ} < i < 180^{\circ}$. This definition satisfies the standard astronomical convention. In this work, we only report the solutions for the inclinations that are between $0^{\circ}$ and $90^{\circ}$, corresponding to face-on and edge-on systems, respectively. However, there are another equally likely solutions of $180 - i$. We also quantify how the error in pulsar mass scales with $\cos i$ using simulated ToAs in section \ref{sec:simulation}, and explain why we were not able to measure the pulsar mass in the other four systems.

\subsection{Measuring the Shapiro delay and constraining pulsar and companion masses}\label{sec:mass_measurement}

We saw significant evidence for Shapiro delay in all 7 systems, but could
only accurately measure the pulsar masses in PSRs~J1125$-$6014, J1614$-$2230 and J1909$-$3744.
For PSRs J0101$-$6422, J1101$-$6424, J1514$-$4946, and J1732$-$5049, it was 
not possible for us to constrain the orthometric parameters significantly, as the uncertainties in the pulsar masses were greater than $2 \, {\rm M_{\odot}}$. Instead, for these systems, we assumed the pulsar mass to be greater than $1.2 \, {\rm M_{\odot}}$ and for linearly spaced $\cos i$ values from $0.01$ to $0.99$, with a resolution of $0.01$, we calculated a series of ($h_{3}$, $\varsigma$) pairs. We then used every pair in the timing model and held them fixed while fitting for other parameters, and recorded a $\chi^2$ value from \textsc{tempo2}. Afterwards, we differentiated the $\chi^{2}$ values from the global minimum, $\chi^{2}_{min}$, to find a $\Delta\chi^{2}$ value for every $\cos i$. Finally, we mapped them to a Bayesian likelihood function \citep{snac02}, and calculated the $1\sigma$ standard deviation of $\cos i$. This gave an estimate of the orbital inclination angle, and using the mass function formula (Eq. \ref{eq:mass_func}), we could find the $1\sigma$ standard deviation of the companion mass. We repeated this analysis for a heavy pulsar mass of $2.0 \, {\rm M_{\odot}}$ as well. Using the two, we could determine how likely the companion mass and the orbital inclination angle of the systems are, assuming the pulsar mass to be between $1.2 \, {\rm M_{\odot}}$ and $2.0 \, {\rm M_{\odot}}$. We believe that the chosen values of $1.2 \, {\rm M_{\odot}}$ and $2.0 \, {\rm M_{\odot}}$ are justifiable as almost all known MSPs lie in this range, and we use this as a prior to explore what the companion masses and inclination angles are. These reported upper and lower limits of $M_{\rm c}$ and $i$ are therefore reasonably conservative. Recently, it has been argued that PSR~J0952$-$0607 has a pulsar mass of $2.35 \pm 0.17 \, {\rm M_{\odot}}$ by examining its luminosity as a function of orbital phase and using its mass ratio from spectroscopic observations of the companion \citep{rkf+22}. If MSPs can be as massive as $\sim 2.35 \, {\rm M_{\odot}}$ this would weaken these companion mass and inclination angle constraints but only mildly.

\subsubsection{PSR~J0101$-$6422}
PSR~J0101$-$6422 is an MSP with a spin period of $2.57 \, \rm ms$ in a $1.78 \, \rm d$ orbit. It is an unidentified $\gamma$-ray source in the first \textit{Fermi} Large Area Telescope catalog \citep{aaa+10}, and was confirmed as an MSP with the Parkes radio telescope, also known as Murriyang, by \cite{kcj+12}. The results from the noise analysis showed that the model containing the white noise and the DM noise parameters had the highest Bayesian evidence. Comparing two timing models including one with the Shapiro parameters and one without, resulted in obtaining a log Bayes factor of $\ln{\mathcal{B}} \sim 21$, demonstrating that the model including the Shapiro parameters is preferred. We detected the Shapiro delay with $\sim 4\sigma$ significance in $h_3$ and $\sim 8\sigma$ significance in $\varsigma$. Using the $\chi^{2}$ analysis mentioned in \ref{sec:mass_measurement}, we place a constraint on $\cos i$ to be in the range of $0.24 \leq \cos i \leq 0.28$ if $M_{\rm p} = 1.2 \, {\rm M_{\odot}}$, and $0.29 \leq \cos i \leq 0.34$ if $M_{\rm p} = 2.0 \, {\rm M_{\odot}}$. The corresponding companion mass and orbital inclination angle for the pulsar mass of $M_{\rm p} = 1.2 \, {\rm M_{\odot}}$ found to be in the ranges of $0.14 \, {\rm M_{\odot}} \leq M_{\rm c} \leq 0.16 \, {\rm M_{\odot}}$ and $73^\circ \leq i \leq 75^\circ$, and for the pulsar mass of $M_{\rm p} = 2.0 \, {\rm M_{\odot}}$ found to be in the ranges of $0.20 \, {\rm M_{\odot}} \leq M_{\rm c} \leq 0.22 \, {\rm M_{\odot}}$ and $70^\circ \leq i \leq 74^\circ$, respectively. These values are consistent with the predictions of \cite{ts99} for a Helium white dwarf (He WD) star. Although these values are reasonably well bounded, they do not lead to interesting constraints on the pulsar's mass.

\subsubsection{PSR~J1101$-$6424}
PSR~J1101$-$6424 is an MSP with a spin period of $5.11 \, \rm ms$ in a $9.61 \, \rm d$ orbit. \cite{ncb+15} discovered the pulsar, during processing of the low-latitude Galactic plane data sets in the High Time Resolution Universe pulsar survey. Comparing the noise models, the model with the white noise and the DM noise parameters had the highest Bayesian evidence. We performed the noise analyses using two timing models with and without the Shapiro delay parameters, and obtained a log Bayes factor of $\ln{\mathcal{B}} \sim 5$. This indicates that we have a weak detection of Shapiro delay in our timing residuals. Using the $\chi^{2}$ analysis mentioned in \ref{sec:mass_measurement}, we place a constraint on $\cos i$ to be in the range of $0.47 \leq \cos i \leq 0.60$ if $M_{\rm p} = 1.2 \, {\rm M_{\odot}}$, and $0.55 \leq \cos i \leq 0.67$ if $M_{\rm p} = 2.0 \, {\rm M_{\odot}}$. The corresponding companion mass and orbital inclination angle for the pulsar mass of $M_{\rm p} = 1.2 \, {\rm M_{\odot}}$ found to be in the ranges of $0.51 \, {\rm M_{\odot}} \leq M_{\rm c} \leq 0.59 \, {\rm M_{\odot}}$ and $53^\circ \leq i \leq 62^\circ$, and for the pulsar mass of $M_{\rm p} = 2.0 \, {\rm M_{\odot}}$ found to be in the ranges of $0.74 \, {\rm M_{\odot}} \leq M_{\rm c} \leq 0.87 \, {\rm M_{\odot}}$ and $47^\circ \leq i \leq 57^\circ$, respectively.
This is an interesting system, as the spin period (and even the orbital period) would suggest it is similar to J1614$-$2230, meaning that is possibly originated as Case A RLO (see discussion in section~\ref{sec:discussion}). The value of $M_{\rm c}$ might be consistent with that of J1614$-$2230.

\subsubsection{PSR~J1125$-$6014}
PSR~J1125$-$6014 is a MSP with a spin period of $2.63 \, \rm ms$ in a $8.75 \, \rm d$ orbit. It is one of the 15 MSPs discovered by \cite{fsk+04} through a reprocessing of the Parkes Multibeam Pulsar Survey. The results from performing noise analysis on the PSR~J1125$-$6014 timing residuals shows that the noise model with the white noise and the DM noise parameters had the highest Bayesian evidence. We derive and constrain the mass of the pulsar to be $M_{\rm p} = 1.68 \pm 0.17 \, {\rm M_{\odot}}$. The companion mass and the orbital inclination angle are constrained to be $M_{\rm c} = 0.33 \pm 0.02 \, {\rm M_{\odot}}$ and $i = 77.6 \pm 0.8^\circ$, respectively. In Figure \ref{fig:J1125_PDF} of Appendix \ref{sec:probability_distributions}, the posterior probability distributions of the measured Shapiro parameters are presented as well as the derived masses and inclination angles for PSR~J1125$-$6014. We discuss the nature of this system in more detail in Section~\ref{sec:discussion}.

\subsubsection{PSR~J1514$-$4946}
PSR~J1514$-$4946 is an MSP with a spin period of $3.59 \, \rm ms$ in a $1.92 \, \rm d$ orbit. It is an unknown source in the first \textit{Fermi} Large Area Telescope catalog \citep{aaa+10}, and was confirmed as an MSP with the Parkes radio telescope by \cite{kcj+12}. Its phase-connected timing solution was later determined by \cite{ckr+15}. We performed a noise analysis on the PSR~J1514$-$4946 timing residuals and obtained the same Bayesian evidences for noise models with DM, red, and white noise parameters. Therefore, we chose the noise model contained only the white noise parameters as it is the simplest one. We obtained a log Bayes factor of $\ln{\mathcal{B}} \sim 6$ by comparing the timing models with and without the Shapiro parameters, showing that we detected a weak Shapiro signature in the timing residuals. Using the $\chi^{2}$ analysis aforementioned in \ref{sec:mass_measurement}, we place a constraint on $\cos i$ to be in the range of $0.14 \leq \cos i \leq 0.28$ if $M_{\rm p} = 1.2 \, {\rm M_{\odot}}$, and $0.21 \leq \cos i \leq 0.36$ if $M_{\rm p} = 2.0 \, {\rm M_{\odot}}$. The corresponding companion mass for the pulsar mass of $M_{\rm p} = 1.2 \, {\rm M_{\odot}}$ found to be in the ranges of $0.15 \, {\rm M_{\odot}} \leq M_{\rm c} \leq 0.17 \, {\rm M_{\odot}}$ and $73^\circ \leq i \leq 82^\circ$, and for the pulsar mass of $M_{\rm p} = 2.0 \, {\rm M_{\odot}}$ found to be in the ranges of $0.22 \, {\rm M_{\odot}} \leq M_{\rm c} \leq 0.24 \, {\rm M_{\odot}}$ and $68^\circ \leq i \leq 78^\circ$, respectively. These values of $M_{\rm c}$ are consistent with the \cite{ts99} prediction for a He WD companion.

\subsubsection{PSR~J1614$-$2230}
\begin{figure}
\centering
 \includegraphics[width=0.5\textwidth]{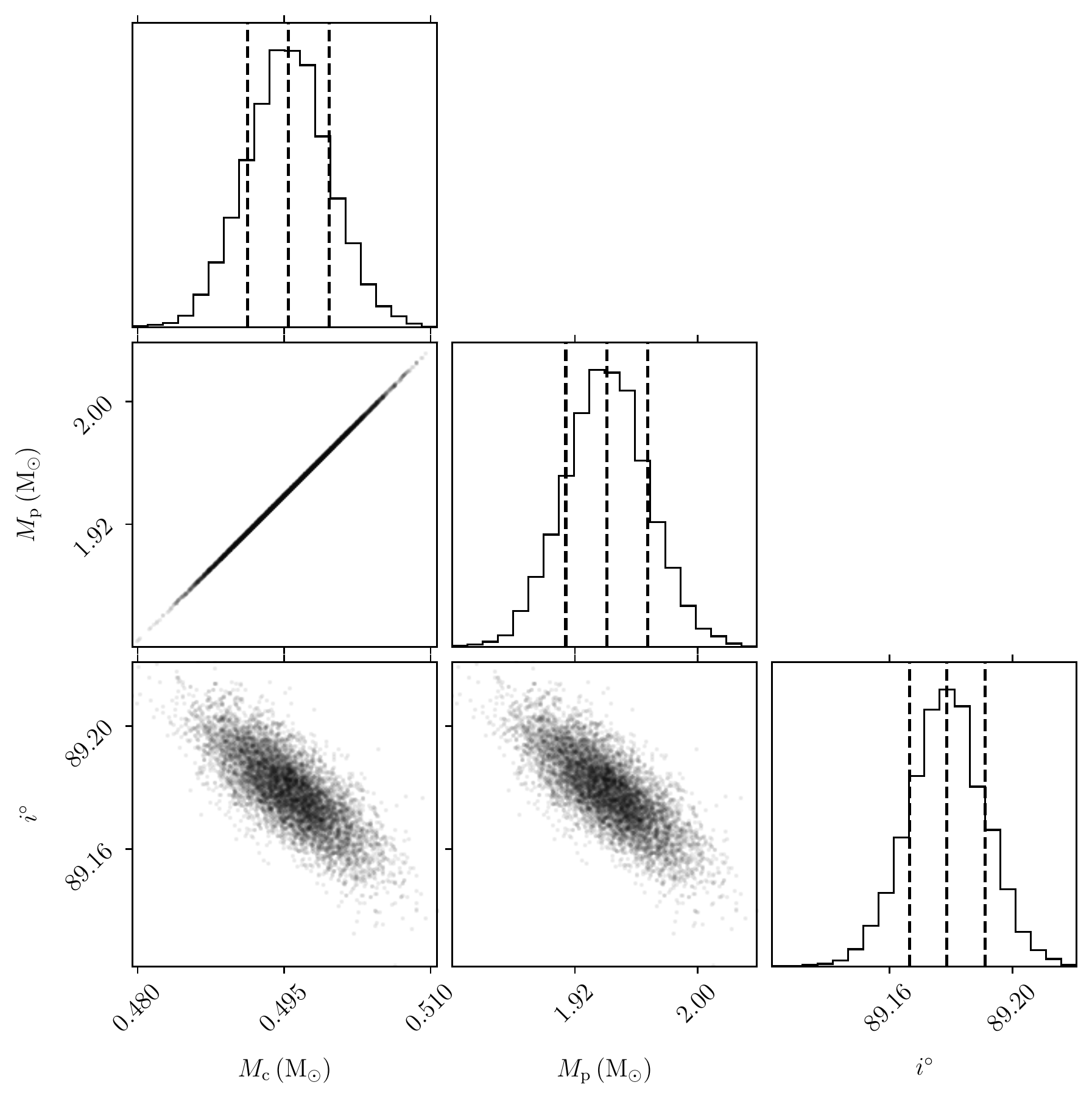}
\caption{2D marginalized posterior probability distributions of $M_{\rm p}$, $M_{\rm c}$, and $i$ (derived from measured $h_{3}$ and $\varsigma$) for J1614$-$2230. }
\label{fig:J1614_pdf_m2_m1_i}
\end{figure}

\begin{figure}
\centering
 \includegraphics[width=0.5\textwidth]{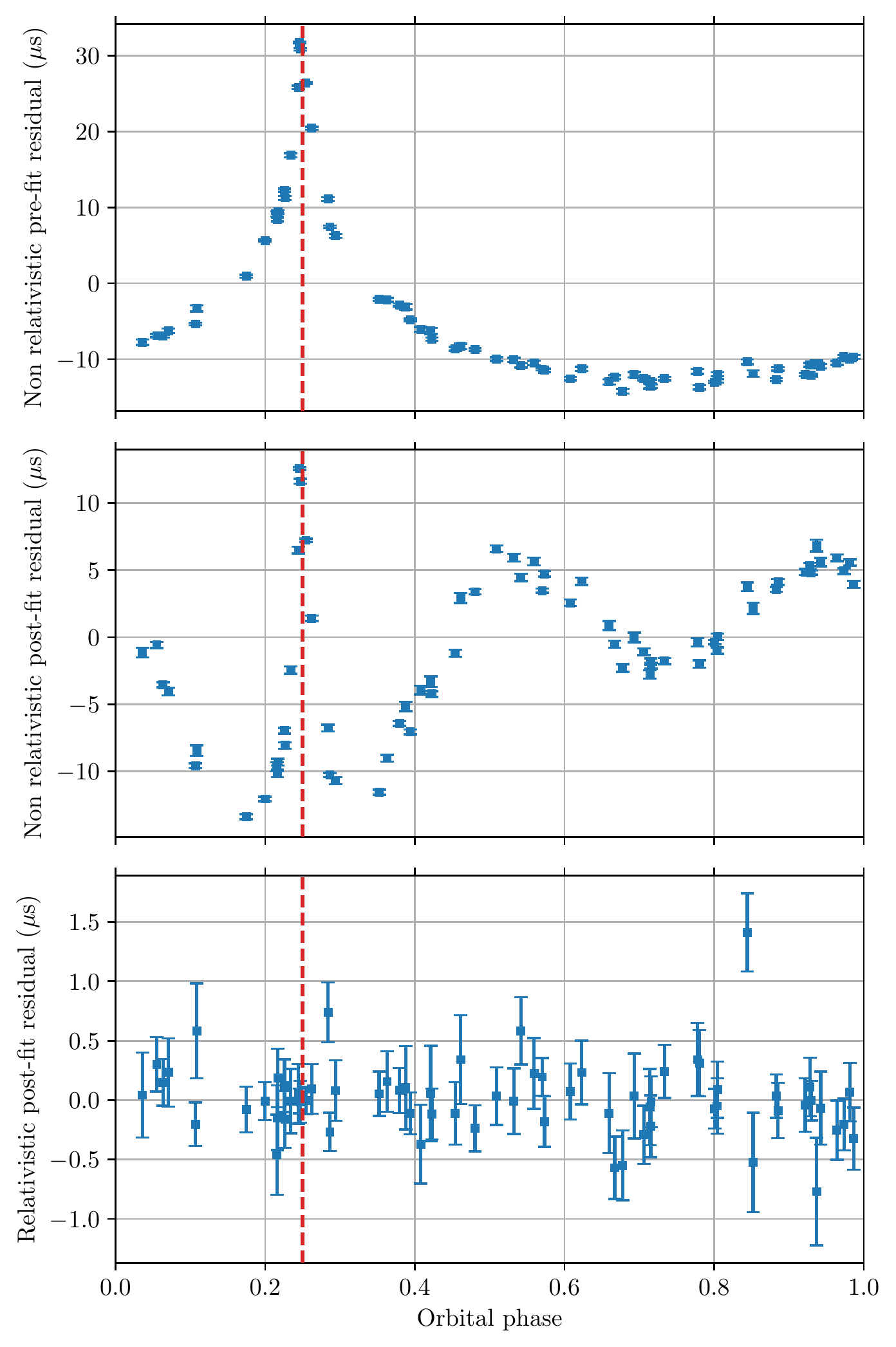}
\caption{Pre- and post-fit timing residuals as a function of orbital phase for J1614$-$2230. The top panel shows the actual Shapiro delay signature preserved in the pre-fit timing residuals (excluding relativistic effects from the timing model). The middle panel shows the timing residuals after fitting for the non-relativistic orbital parameters. The bottom panel show the timing residuals after fitting for all the parameters. The superior conjunction happens at the orbital phase of 0.25 and is indicated by the vertical dashed red line.}
\label{fig:J1614}
\end{figure}

 PSR~J1614$-$2230 is an MSP with a spin period of $3.15 \, \rm ms$ in a $8.69 \, \rm d$ orbit. \cite{crh+06a} discovered PSR~J1614$-$2230 after conducting an intermediate-latitude Galactic survey of 56 unidentified $\gamma$-ray sources from the third Energetic Gamma-Ray Experiment Telescope catalog \citep{hbb+99} with the 13 beam Parkes multibeam receiver at $1400 \rm MHz$ \citep{hrr+05,crh+06a}. The precise Shapiro delay measurement, and high mass derived for PSR~J1614$-$2230 carried out by \cite{dpr+10} was able to reject many of the EOSs of the nuclear matter. Among our pulsars, PSR~J1614$-$2230 has the most significant Shapiro delay detection: $h_3 = 2.340 \pm 0.019 \, \mu$s and $\varsigma = 0.9858 \pm 0.0003$. These led us to constrain the mass of the pulsar to be $M_{\rm p}=1.94 \pm 0.03 \, {\rm M_{\odot}}$, the companion mass to be $M_{\rm c}=0.495 \pm 0.005 \, {\rm M_{\odot}}$, and the orbital inclination angle to be $i = 89.179^{\circ} \pm 0.013^{\circ}$. Figure \ref{fig:J1614_pdf_m2_m1_i} shows the correlations between these parameters. The timing residuals, pre- and post-fitting, are shown in Figure \ref{fig:J1614}. The posterior probability distributions of $h_3$ and $\varsigma$, as well as $M_{\rm c}$, $M_{\rm p}$, and $i$ that are derived from $h_3$ and $\varsigma$, are plotted in Figure \ref{fig:J1614_PDF} of Appendix \ref{sec:probability_distributions}. 
 For PSR~J1614$-$2230, the noise model containing the DM noise, red noise, and white noise parameters had the highest Bayesian evidence.
 
\subsubsection{PSR~J1732$-$5049}
 PSR~J1732$-$5049 is an MSP with a spin period of $5.31 \, \rm ms$ in a $5.26 \, \rm d$ orbit. It is one of the nine pulsars discovered by \cite{eb01b} in an intermediate-latitude Galactic survey with a centre frequency near $1400 \, \rm MHz$ conducted with the Parkes radio telescope. From our timing noise analysis, we obtained the highest evidence for the noise model containing the red noise and white noise parameters. Comparing the timing solutions with and without the Shapiro parameters, PSR~J1732$-$5049 showed a likely Bayesian evidence for the presence of a Shapiro signature with a log Bayes factor of $\log\mathcal{B} \sim 6$. Using the $\chi^{2}$ analysis mentioned in \ref{sec:mass_measurement}, we place a constraint on $\cos i$ to be in the range of $0.31 \leq \cos i \leq 0.47$ if $M_{\rm p} = 1.2 \, {\rm M_{\odot}}$, and $0.35 \leq \cos i \leq 0.51$ if $M_{\rm p} = 2.0 \, {\rm M_{\odot}}$. The corresponding companion mass for the pulsar mass of $M_{\rm p} = 1.2 \, {\rm M_{\odot}}$ was found to be in the ranges of $0.17 \, {\rm M_{\odot}} \leq M_{\rm c} \leq 0.20 \, {\rm M_{\odot}}$ and $61^\circ \leq i \leq 72^\circ$, and for the pulsar mass of $M_{\rm p} = 2.0 \, {\rm M_{\odot}}$ found to be in the ranges of $0.24 \, {\rm M_{\odot}} \leq M_{\rm c} \leq 0.27 \, {\rm M_{\odot}}$ and $59^\circ \leq i \leq 70^\circ$, respectively.
 These values of $M_{\rm c}$ are consistent with the \cite{ts99} prediction for a He WD companion.

\subsubsection{PSR~J1909$-$3744}
PSR~J1909$-$3744 is an MSP with a spin period of $2.95 \, \rm ms$ in a $1.53 \, \rm d$ orbit. The Swinburne High Latitude Pulsar Survey was a 
blind survey that led \cite{jbk+03} to discover PSR~J1909$-$3744 using the Parkes 13-beam multibeam receiver. The noise model including DM noise, red noise, and white noise parameters had the highest evidence, and our constraints on the orthometric Shapiro parameters were $h_3 = 0.845 \pm 0.007 \, \mu$s and $\varsigma = 0.9441 \pm 0.0016$. We derive and constrain the mass of the pulsar to be $M_{\rm p} = 1.45 \pm 0.03 \, {\rm M_{\odot}}$. The companion mass and the orbital inclination angle are constrained to be $M_{\rm c} = 0.205 \pm 0.003 \, {\rm M_{\odot}}$ and $i = 86.69^\circ \pm 0.10$, respectively. The posterior probability distribution of the measured and derived parameters can be found in the Figure \ref{fig:J1909_PDF} of Appendix \ref{sec:probability_distributions}.  The companion mass is consistent with the predictions of \cite{ts99} for a He WD, and is the first confirmation of this mass relation.

\begin{table}
\centering
\caption{Pulsar and companion star masses and inclination angles. 
The derived inclination angles have another equally likely solution of $180^\circ - \, i$. 
The pulsars with two entries could not have their pulsar masses well determined and so
two possible (small and large) pulsar masses were held fixed and likely companion and inclination angles
provided for them.}
\begin{tabular}{@{}lllll}
\hline
Pulsar & $M_{\rm p}$ ($\rm M_{\odot}$) & $M_{\rm c}$ ($\rm M_{\odot}$) & $\cos i$ & $i^{\circ}$  \\
\hline
J0101$-$6422  & $1.2$ & $0.14$--$0.16$ & $0.24$--$0.28$ & $73$--$75$ \\
  & $2.0$ & $0.20$--$0.22$ & $0.29$--$0.34$ & $70$--$74$ \\

J1101$-$6424  & $1.2$ & $0.51$--$0.59$ & $0.47$--$0.60$ & $53$--$62$ \\
  & $2.0$ & $0.74$--$0.87$ & $0.55$--$0.67$ & $47$--$57$ \\

J1125$-$6014  & $1.68^{+0.17}_{-0.15}$ & $0.33^{+0.02}_{-0.02}$ & $0.215^{+0.013}_{-0.012}$ & $77.6^{+0.7}_{-0.8}$ \\

J1514$-$4946  & $1.2$ & $0.15$--$0.17$ & $0.14$--$0.28$ & $73$--$82$ \\
  & $2.0$ & $0.22$--$0.24$ & $0.21$--$0.36$ & $68$--$78$ \\

J1614$-$2230 & $1.94^{+0.03}_{-0.03}$ & $0.495^{+0.005}_{-0.005}$ & $0.0143^{+0.0002}_{-0.0002}$ & $89.179^{+0.013}_{-0.013}$ \\

J1732$-$5049  & $1.2$ & $0.17$--$0.20$ & $0.31$--$0.47$ & $61$--$72$ \\
  & $2.0$ & $0.24$--$0.27$ & $0.35$--$0.51$ & $59$--$70$ \\

J1909$-$3744 & 1.45$^{+0.03}_{-0.03}$ & $0.205^{+0.003}_{-0.003}$ & $0.0573^{+0.0016}_{-0.0016}$ & $86.69^{+0.10}_{-0.10}$ \\
\hline
\end{tabular}
\label{tab:mass_inclination}
\end{table}

\subsection{The uncertainty in the pulsar mass as a function of cos(i)}\label{sec:simulation}

While it is unfortunate that not all pulsar binaries are in edge-on systems, it is not surprising. Here we explore the $(M_{\rm c}, \sin i)$ space where meaningful Shapiro delays can be detected through simulation. For an arbitrary binary that we simulate, we wanted to understand what the companion mass error ($\sigma_{\rm M_{c}}$) and $\sin i$ error ($\sigma_{\rm \sin i}$) are by holding the pulsar mass, companion mass, and orbital period fixed, and then use them to understand what the pulsar mass error ($\sigma_{\rm M_{\rm p}}$) would be after the fit. 

Using \textsc{tempo2}'s \texttt{fake} plugin, we conducted two sets of simulations. In both simulations, we have used ELL1
binary model. In the first, we investigated how $\sigma_{\rm M_{c}}$ and $\sigma_{\rm \sin i}$ generally change as a function of $\cos i$. Then we used equations \ref{eq:mass_func} and \ref{eq:mp_error} in order to predict $\sigma_{\rm M_{\rm p}}$. The results of this simulation provides a mathematical relation by which the error in the pulsar mass $\sigma_{\rm M_{\rm p}}$ can be estimated. 

\subsubsection{Constant M$_{\rm c}$ and M$_{\rm p}$}\label{sec:sim_the_same_system}
\begin{figure}
\centering
 \includegraphics[width=0.5\textwidth]{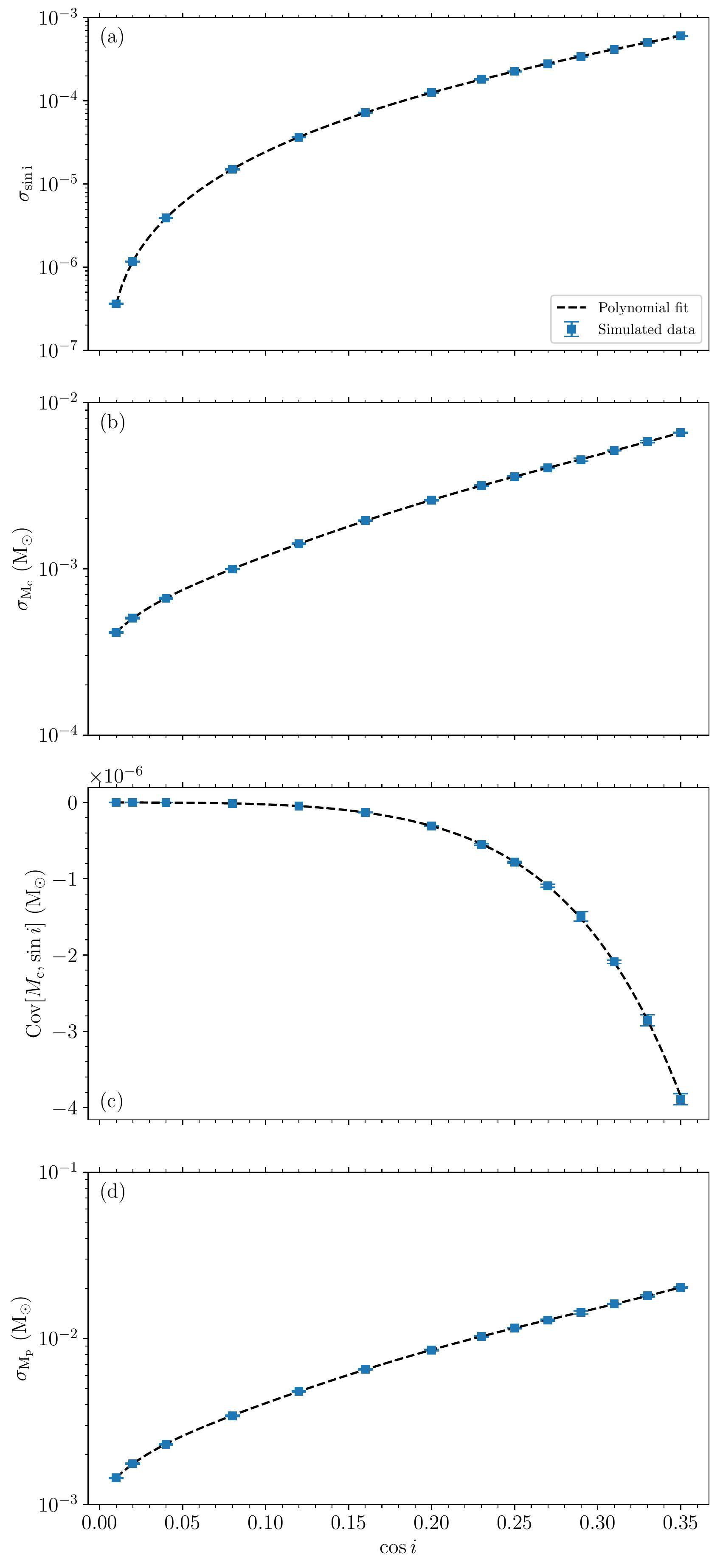}
\caption{Exploring the error in the component masses at different orbital inclination angles using $1300$ simulated ToAs with the rms residuals of $0.1 \, \mu$s, for a binary pulsar system with the companion mass of $1 \, {\rm M_{\odot}}$, the pulsar mass of $2.0 \, {\rm M_{\odot}}$, and the orbital period of $8.69$ d. (a): the error in the sine of the orbital inclination angle. (b): the error in the companion mass. (c): the covariance between the companion mass and the sine of the orbital inclination angle. (d): the error in the pulsar mass. The dashed lines in panels (a), (b), and (c) are the polynomial functions described by Equations \ref{eq:error_model_sini}, \ref{eq:error_model_mc}, \ref{eq:error_model_covariance}, respectively, and the coefficients for each are listed in Table \ref{tab:error_model_results}. The dashed line in the bottom panel is calculated using Equation \ref{eq:mp_error}.}
\label{fig:J1614_dmp_dcosi_same_system}
\end{figure}

In the first set of simulations, we aimed to perform the analysis of measuring $\sigma_{\rm M_{\rm p}}$ as we are rotating the orbit in a way that the inclination angle of the orbit changes with respect to our line of sight. We took our timing solution for J1614$-$2230, removed the dispersion measure and the noise parameters, and used it to simulate ToAs. Using Equation \ref{eq:mass_func}, the projected semi-major axis of the pulsar at different inclination angles from $45^\circ$ to $89^\circ$, given the pulsar mass of $2.0 \, {\rm M_{\odot}}$ and the companion mass of $1.0 \, {\rm M_{\odot}}$, were calculated. Then $1300$ ToAs with the root-mean-square (rms) residuals of $ 0.1 \, \mu$s were simulated
(similar to what can be obtained for some MSPs in a reasonable time), and using \textsc{temponest} we sampled the binary parameters including $P_{\rm b}$, $T_{\rm asc}$, $x$, $\epsilon_{1}$, $\epsilon_{2}$, $M_{\rm c}$, and $\sin i$ and recorded the values and errors of $\sin i$,  $M_{\rm c}$, and $\operatorname{Cov}[M_{\rm c},\sin i]$. For every inclination angle, the simulation was repeated $10$ times. In panel (a) of Figure \ref{fig:J1614_dmp_dcosi_same_system}, the value and the error bar of every point is, respectively, the average and the $1\sigma$ standard deviation of $\sigma_{\rm \sin i}$ taken from $10$ \textsc{temponest} runs. The same procedure was used for $\sigma_{\rm M_{\rm c}}$, $\operatorname{Cov}[M_{\rm c},\sin i]$, and $\sigma_{\rm M_{\rm p}}$ in the panels (a), (b), (c), respectively. For every \textsc{temponest} run, $\sigma_{\rm M_{\rm p}}$ was calculated using Equation\ref{eq:mp_error}.

We next derived empirical relationships between $\cos i$ and $\sigma_{\rm \sin i}$, $\sigma_{\rm M_{c}}$, and $\operatorname{Cov}[M_{\rm c},\sin i]$.
We found that the best fitting function to the relationships was of the polynomial form
\begin{equation}\label{eq:error_model_formula}
    \mathcal{F}(\cos i) = \sum_{\rm j=0}^{n} a_{\rm j} \cos ^{\rm j} i,
\end{equation}
where the coefficients of $a_{\rm j}$ for $\sigma_{\rm M_{c}}$ fitting function  ($\mathcal{F}_{\rm \sigma_{M_{c}}}$), $\sigma_{\rm \sin i}$ fitting function ($\mathcal{F}_{\rm \sigma_{\sin i}}$), and $\operatorname{Cov}[M_{\rm c},\sin i]$ fitting function ($\mathcal{F}_{\operatorname{Cov}[M_{\rm c},\sin i]}$) can be found by minimizing the sum of least squares. The number of coefficients depends on the orbital inclination angle. We found that, for $0.01 \leq \cos i \leq 0.16$, we need at least a third order polynomial ($n=3$). By increasing the range of $\cos i$ to be in $0.01 \leq \cos i \leq 0.35$, corresponding to the orbital inclination angle of $70^\circ \leq i < 90^\circ$, we need to have at least six coefficients. We used the latter case in our modeling of uncertainties, and found the constant coefficients of $a_{0},..., a_{5}$ in Table \ref{tab:error_model_results} for $\mathcal{F}_{\rm \sigma_{\sin i}}$ (the dashed line in the panel (a) of Figure \ref{fig:J1614_dmp_dcosi_same_system}), $\mathcal{F}_{\rm \sigma_{M_{c}}}$ (the dashed line in the panel (b) of Figure \ref{fig:J1614_dmp_dcosi_same_system}), and $\mathcal{F}_{\operatorname{Cov}[M_{\rm c},\sin i]}$ (the dashed line in the panel (c) of Figure \ref{fig:J1614_dmp_dcosi_same_system}). We then used Equation \ref{eq:mp_error} and substituted $\sigma_{\rm M_{c}}$, $\sigma_{\rm \sin i}$, and $\operatorname{Cov}[M_{\rm c},\sin i]$ with $\mathcal{F}_{\rm \sigma_{M_{c}}}$, $\mathcal{F}_{\rm \sigma_{\sin i}}$, and $\mathcal{F}_{\operatorname{Cov}[M_{\rm c},\sin i]}$, respectively, to find the modeled error in the pulsar mass (the dashed line in the panel (d) of Figure \ref{fig:J1614_dmp_dcosi_same_system}). We also explored how $\sigma_{\rm \sin i}$, $\sigma_{\rm M_{c}}$, and $\operatorname{Cov}[M_{\rm c},\sin i]$ scale with the rms residuals of $1$ and $2 \, \mu$s, the companion masses of $0.2$ and $0.5 \, {\rm M_{\odot}}$, and with $2600$ ToAs (2 times the initial number of ToAs). The results show that $\sigma_{\rm \sin i}$, $\sigma_{\rm M_{c}}$, and $\operatorname{Cov}[M_{\rm c},\sin i]$ are scaled with the following relations:

\begin{equation}\label{eq:error_model_sini}
    \sigma_{\rm \sin i} =  \left ( \frac{{\rm M_{\odot}}}{M_{\rm c}} \right ) \left ( \frac{\sigma_{\rm rms}}{\mu s} \right ) \left ( \frac{1000}{N_{\rm ToA}} \right )^{\frac{1}{2}} \mathcal{F}_{\rm \sigma_{\sin i}}(\cos i),
\end{equation}
\begin{equation}\label{eq:error_model_mc}
    \sigma_{\rm M_{c}} = \left ( \frac{\sigma_{\rm rms}}{\mu s} \right ) \left ( \frac{1000}{N_{\rm ToA}} \right )^{\frac{1}{2}} \mathcal{F}_{\rm \sigma_{M_{c}}}(\cos i),
\end{equation}
and
\begin{equation}\label{eq:error_model_covariance}
    \operatorname{Cov}[M_{\rm c},\sin i] =  \left ( \frac{{\rm M_{\odot}}}{M_{\rm c}} \right ) \left ( \frac{\sigma_{\rm rms}}{\mu s} \right )^{2} \left ( \frac{1000}{N_{\rm ToA}} \right ) \mathcal{F}_{\rm \operatorname{Cov}[M_{\rm c},\sin i]}(\cos i).
\end{equation}

\begin{table}
\centering
\caption{Fitting results for the error modelling.}
\begin{tabular}{llll}
\hline
coefficient & $\mathcal{F}_{\rm \sigma_{\sin i}}$ & $\mathcal{F}_{\rm \sigma_{M_{c}}}$ & $\mathcal{F}_{\rm \operatorname{Cov}[M_{\rm c},\sin i]}$ \\
\hline
$a_{0}$ & $-1.7(4) \times 10^{-6}$ & $0.00363(10)$ & $-9(11) \times 10^{-9}$\\
$a_{1}$ & $4.3(6) \times 10^{-4}$ & $0.116(9)$ & $2(2) \times 10^{-6}$ \\
$a_{2}$ & $0.014(2)$ & $-0.7(2)$ & $-2.6(11) \times 10^{-4}$ \\
$a_{3}$ & $0.10(3)$ & $7.0(19)$ & $3(3)$ \\
$a_{4}$ & $0.10(15)$ & $-22(7)$ & $-0.05(2)$ \\
$a_{5}$ & $0.4(2)$ & $27(8)$ & $0.17(9)$ \\
$a_{6}$ & $-$ & $-$ & $-0.38(13)$ \\
\hline
\end{tabular}
\label{tab:error_model_results}
\end{table}

We used our model to calculate $\sigma_{\rm M_{p}}$ for both PSRs~J1614$-$2230 and J1909$-$3744. Because the ToA uncertainties range from $201 \, \rm ns$ to $7.6 \, \mu \rm s$ for PSR~J1614$-$2230, and from $9 \, \rm ns$ to $2.2 \, \mu \rm s$ for PSR~J1909$-$3744, we needed to create ToAs that are averaged in time. We averaged to $4 \, \rm min$ (the observing length of typical observations), and then calculated the number of unique averaged ToAs and the weighted rms of the residuals. After averaging, for PSR~J1614$-$2230 we obtained a weighted rms residual of $0.310 \, \mu \rm s$ for $119$ observations, and for PSR~J1909$-$3744 we obtained a weighted rms residual of $0.125 \, \mu \rm s$ for $179$ observations. Using our model, we determined $\sigma_{\rm M_{p}}$ to be $\sim 0.03 \, {\rm M_{\odot}}$ for both, which is in very good agreement with the measured $\sigma_{\rm M_{p}}$ of $0.03 \, {\rm M_{\odot}}$ in Table \ref{tab:mass_inclination}. 

We are able to use our model to estimate $\sigma_{\rm M_{p}}$ in PSR~J0101$-$6422 and J1514$-$4946, but not in PSRs~J1101$-$6424 and J1732$-$5049, because our model is only reliable for $\cos i < 0.35$. For PSR~J0101$-$6422 and J1514$-$4946, if we assume $M_{\rm p} = 1.4 \, {\rm M_{\odot}}$, by taking the median values of $\cos i$ to be $0.27$ and $0.22$, resulting in the companion masses to be $0.17 \, {\rm M_{\odot}}$ and $0.18 \, {\rm M_{\odot}}$, we estimated $\sigma_{\rm M_{p}}$ to be $1.6 \, {\rm M_{\odot}}$ and $0.8 \, {\rm M_{\odot}}$, respectively. If we assume $M_{\rm p} = 2.0 \, {\rm M_{\odot}}$, by taking the median values of $\cos i$ to be $0.30$ and $0.27$, resulting in the companion masses to be $0.21 \, {\rm M_{\odot}}$ and $0.23 \, {\rm M_{\odot}}$, we estimated $\sigma_{\rm M_{p}}$ to be $2.0 \, {\rm M_{\odot}}$ and $1.3 \, {\rm M_{\odot}}$, respectively. 

\subsubsection{Constant $f_{\rm m}$}
Soon after a binary pulsar is discovered, the mass function is quickly (and accurately) established, and so we also wanted to explore what the likely $\sigma_{\rm M_{\rm p}}$ would be if the pulsar mass and mass function are held fixed and we have no a priori knowledge of the
inclination angle. This simulation method can be used whenever a new 
millisecond pulsar is discovered, the mass function is known, and
observers want to determine the chance of being able to measure a significant Shapiro delay
for various inclination angles (and hence component masses). 
In this second set of simulations, we thus investigated how $\sigma_{\rm M_{\rm p}}$ depended upon the known mass functions in PSRs~J0101$-$6422, J1101$-$6424, J1514$-$4946, and J1732$-$5049 if $80^\circ \leq i < 90^\circ$, corresponding to $\sim 0.0 \leq \cos i \leq 0.16$. We did not explore higher $\cos i$ values as beyond this $\sigma_{\rm M_{p}}$ becomes larger than $2.0 \, {\rm M_{\odot}}$. The companion mass values were calculated at different inclination angles using Equation \ref{eq:mass_func}, assuming the mass function of each pulsar and $M_{\rm p} = 1.5 \, {\rm M_{\odot}}$. We then simulated $1300$ ToAs with the rms residuals of data in the Tables \ref{tab:timing_solution_1} and \ref{tab:timing_solution_2}. The same procedure as Section \ref{sec:sim_the_same_system} was used for performing the timing analysis, and plotting the errors. The only difference was, at every inclination angle, the simulation was repeated 100 times. 

\begin{figure}
\centering
 \includegraphics[width=0.5\textwidth]{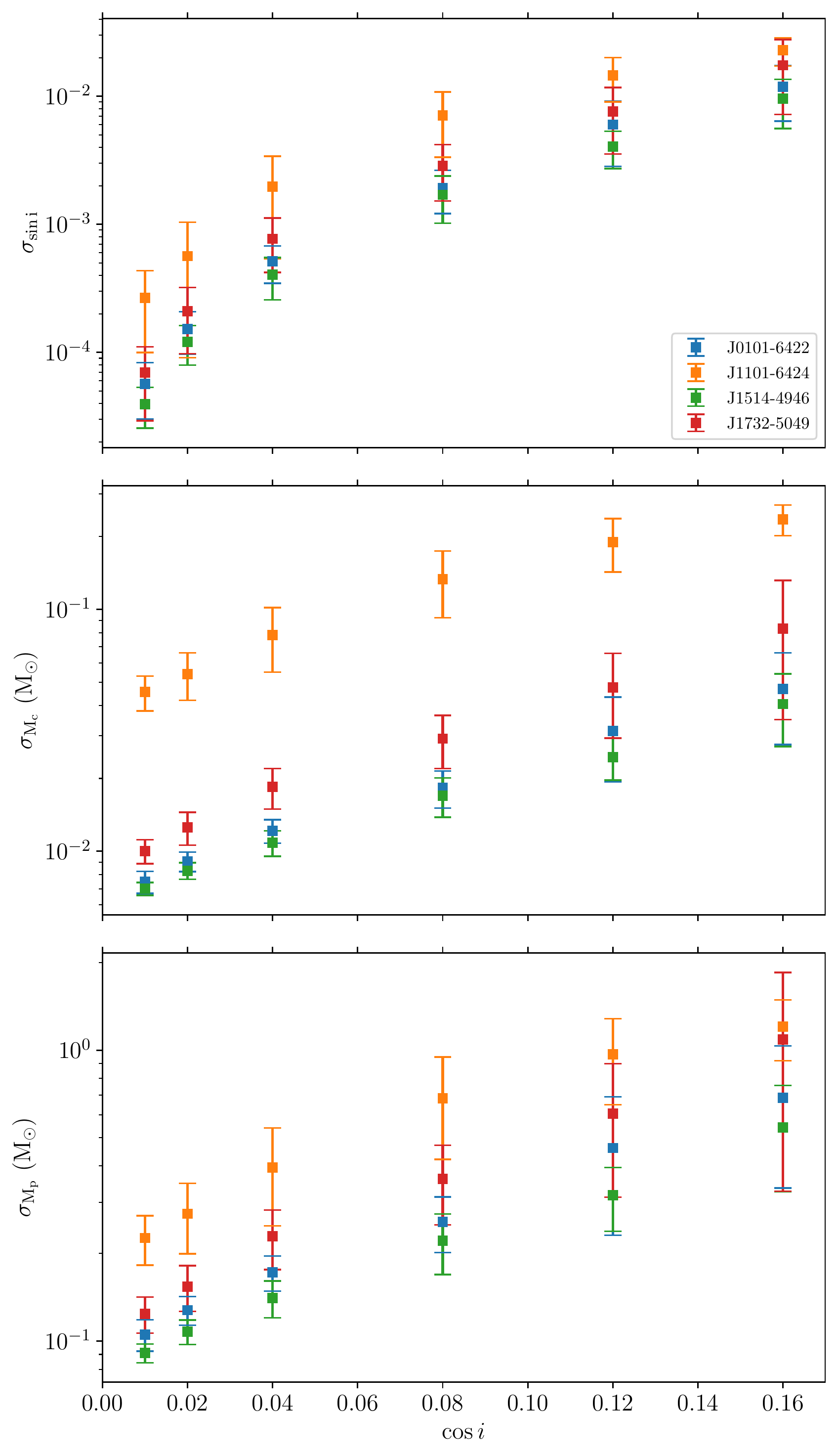}
\caption{Error in the component masses at different orbital inclination angles for PSRs~J0101$-$6422, J1101$-$6424, J1514$-$4946, and J1732$-$5049 using 1300 simulated ToAs with the rms of the actual data. Top: the error in the sine of orbital inclination angle. Middle: the error in the companion mass. Bottom: the error in the pulsar mass. The colors indicate the different binary pulsar systems of PSRs~J0101$-$6422(blue), J1101$-$6424 (orange), J1514$-$4946 (green), and J1732$-$5049 (red). 
The mass of the pulsar is assumed to be $1.5 \, {\rm M_{\odot}}$ in all the simulations.}
\label{fig:dmp_dcosi_real_data}
\end{figure}

In Figure \ref{fig:dmp_dcosi_real_data}, PSR~J1101$-$6424 has the largest uncertainties in $\sin i$ and $M_{\rm c}$ due to its high rms residuals of $7.973 \, \mu$s despite its massive companion. In our timing analysis for PSRs~J0101$-$6422, J1101$-$6424, J1514$-$4946, and J1732$-$5049, we were interested in getting $\sigma_{\rm M_{p}} \leq 0.1 \, {\rm M_{\odot}}$ to be astrophysically interesting. However, from Figure \ref{fig:dmp_dcosi_real_data}, the orbital inclination angles of PSRs~J0101$-$6422 and J1514$-$4946 systems need to have $\cos i \leq 0.01$ to achieve this, given our timing precision. PSRs~J1101$-$6424 and J1732$-$5049 have $\sigma_{\rm M_{p}}$ greater than $0.1 {\rm M_{\odot}}$ at all values of $\cos i$. From the $\chi^{2}$ analysis in \ref{sec:mass_measurement}, all four pulsars have $\cos i$ greater than $0.16$, which stops us from placing meaningful constraints on their pulsar masses.

\section{Discussion}\label{sec:discussion}

\subsection{Comparison with previous measurements}

We now check whether our results are consistent with those obtained by other authors, in a historical sequence.

\subsubsection{PSR~J1909$-$3744}

From the NANOGrav $11$ yr data release, \cite{abb+18} measured the Shapiro delay
using the traditional parameterisation ($M_{\rm c} , \sin i$)
for PSR~J1909$-$3744 and estimated the parameters of $M_{\rm p} = 1.48 \pm 0.03 \, {\rm M_{\odot}}$, $M_{\rm c} = 0.208 \pm 0.002 \, {\rm M_{\odot}}$, and $i = 86.47^\circ \pm 0.10^\circ$. \cite{lgi+20} used 15 yr of PSR~J1909$-$3744 observations with Nancay Radio Telescope to perform a high-precision analysis, and derived a pulsar mass of $M_{\rm p} = 1.492 \pm 0.014 \, {\rm M_{\odot}}$ and a companion mass of $M_{\rm c} = 0.209 \pm 0.001 \, {\rm M_{\odot}}$ using the traditional parameterisation. Another independent analysis of J1909$-$3744 used $15$ yr of PPTA-DR2e \citep{rsc+21} to measure the Shapiro delay parameters and obtained $M_{\rm p} = 1.486 \pm 0.011 \, {\rm M_{\odot}}$, $M_{\rm c} = 0.2081 \pm 0.0009 \, {\rm M_{\odot}}$, $i = 86.46^\circ \pm 0.05^\circ$. Using the International Pulsar Timing Array second data release (consisting of data sets collected by PPTA, the European pulsar timing array, and NANOGrav) \citet{pdd+19} analyzed $10.8$ yr of data and provided measurements of the classical Shapiro parameters of $M_{\rm c}=0.209 \pm 0.001 \, {\rm M_{\odot}}$ and $\sin i = 0.99807 \pm 0.00006$ for PSR~J1909$-$3744. Using the mass function from \texttt{psrcat}\footnote{\url{https://www.atnf.csiro.au/research/pulsar/psrcat}} and applying Equations \ref{eq:mass_func} and \ref{eq:mp_error}, we can derive the pulsar mass to be $M_{\rm p} = 1.496 \pm 0.011 \, {\rm M_{\odot}}$. Our derived masses and the inclination angle for PSR~J1909$-$3744 (from the Table \ref{tab:mass_inclination}) are: $M_{\rm p} = 1.45 \pm 0.03 \, {\rm M_{\odot}}$, $M_{\rm c} = 0.205 \pm 0.003 \, {\rm M_{\odot}}$, $i = 86.69^\circ \pm 0.10^\circ$. Our masses are approximately $1\sigma$ smaller than previously reported values.

\subsubsection{PSR~J1614$-$2230}
The analysis of the PSR~J1614$-$2230 system by \cite{dpr+10} utilized a set of observations from the Green Bank Telescope to constrain the Shapiro delay using the classic parameterisation: $M_{\rm p} = 1.97 \pm 0.04 \, {\rm M_{\odot}}$, $M_{\rm c} = 0.500 \pm 0.006 \, {\rm M_{\odot}}$, $i = 89.17^\circ \pm 0.02^\circ$. In a subsequent analysis, \cite{fpe+16} used the NANOGrav 9 yr data set, including a part of the data set used by \cite{dpr+10}, and made significant measurements of the Shapiro delay using both the classic and the orthometric parameterisations: $M_{\rm p} = 1.928 \pm 0.017 \, {\rm M_{\odot}}$, $M_{\rm c} = 0.493 \pm 0.003 \, {\rm M_{\odot}}$, $i = 89.189^\circ \pm 0.014^\circ$. Then \cite{abb+18} used the NANOGrav 11 yr data set and put constraints on the Shapiro delay using the orthometric parameterisation obtaining: $M_{\rm p} = 1.908 \pm 0.016 \, {\rm M_{\odot}}$, $M_{\rm c} = 0.493 \pm 0.003 \, {\rm M_{\odot}}$, $i = 89.204^\circ \pm 0.014^\circ$.

Our reported masses and the inclination angle for PSR~J1614$-$2230 (from the Table \ref{tab:mass_inclination}) are: $M_{\rm p} = 1.94 \pm 0.03 \, {\rm M_{\odot}}$, $M_{\rm c} = 0.495 \pm 0.005 \, {\rm M_{\odot}}$, $i = 89.179^\circ \pm 0.013^\circ$. These are the first independent measurement using MTPA data sets and are in good agreement with the previous measurements reported.

The discovery of PSR~J1614$-$2230 is important for reasons beyond the implications for the study of the EOS of dense matter: it was also important to establish that neutron stars can be born with large masses. \citet{tlk11} carried out a detailed study on modeling the formation and evolution of PSR~J1614$-$2230 based on its physical parameters reported by \citet{dpr+10}. They investigated the evolution of PSR~J1614$-$2230 using an evolving intermediate-mass X-ray binary and reported two possible scenarios: 1) through a common envelope started with their ``Case C'' RLO 2) through the highly super-Eddington isotropic re-emission mode during their ``Case A''  RLO, with the conclusion of the second scenario as the more likely one and the mass of the PSR~J1614$-$2230 progenitor to be higher than $20 {\rm M_{\odot}}$, and the initial companion to be a main-sequence donor star with the mass of $4.0-5.0\, {\rm M_{\odot}}$.

Because in Case A RLO the mass transfer happens while the companion is in the main sequence, the recycling episode is very long; providing a better explanation for the spin period of PSR~J1614$-$2230, which is unusually small compared to all other pulsars with massive WDs. Interestingly, they find that even with the long accretion episode of Case A RLO, the total mass transfer is very small, less that $0.3 \, \rm M_{\odot}$. This means that the birth mass was at least $\sim 1.7 \, \rm M_{\odot}$. This is a strong indication that neutron star masses are generally acquired at birth, and don't owe much to subsequent evolution.

Our measurements confirm their interpretation of the mass of PSR~J1614$-$2230 system. The idea that neutron stars have high masses at birth was later confirmed with greater confidence by the discovery of massive neutron stars where little or no accretion happened, like PSR~J2222$-$0137 \citep{cfg+17,gfg+21}, and further by measurement of the mass of PSR~J1933$-$6211, which is thought to have evolved from case A RLO and nevertheless has a low mass, confirming that not much mass was transferred during its long evolutionary episode (Geyer et al., submitted).

\subsubsection{PSR~J1125$-$6014}

From the the second data release of the Parkes Pulsar Timing Array (PPTA-DR2e), \cite{rsc+21} performed an analysis of the PSR~J1125$-$6014 system using more than $12$ yr of data (a factor of four larger than the time span of PSR~J1125$-$6014 observations with MeerKAT), and they constrained the pulsar mass, the companion mass, and the orbital inclination angle to be $M_{\rm p} = 1.5 \pm 0.2 \, {\rm M_{\odot}}$, $M_{\rm c} = 0.31 \pm 0.03 \, {\rm M_{\odot}}$, $i = 77.9{^\circ} ^{+1.4}_{-1.3}$, respectively. Our derived masses and the inclination angle for PSR~J1125$-$6014 (from the Table \ref{tab:mass_inclination}) are $M_{\rm p} = 1.68^{+0.17}_{-0.15} \, {\rm M_{\odot}}$, $M_{\rm c} = 0.33 \pm 0.02 \, {\rm M_{\odot}}$, $i = 77.6{^\circ} ^{+0.7}_{-0.8}$, which are in good agreement with \cite{rsc+21}.

The mass of the companion of PSR~J1125$-$6014 is significantly higher than the prediction of \cite{ts99} for the orbital period of this system ($0.25\, \rm M_{\odot}$). This suggests that the companion is not likely a Helium WD, but instead a low-mass CO WD. Interestingly, the short spin period of this pulsar and the high degree of recycling also suggest a case A RLO evolution, making the system a less massive version of PSR~J1614$-$2230. We note in this regard the similarity of the orbital periods (both $\sim 8.7\, \rm d$) and orbital eccentricities of the two systems.


\subsection{Mass measurements with Shapiro delay}

As an application of our error model, we can asses the fraction of MSPs for which we expect the masses to be measured to better than $10\%$ using the Shapiro delay method. For a sample of $100$ randomly oriented MSPs, each with a pulsar mass of $1.4 \, {\rm M_{\odot}}$, $1300$ ToAs, and the rms residuals of $1 \, \mu$s, we would expect to measure approximately $11$ MSP masses to better than $10\%$, assuming the companion masses are all $0.2 \, {\rm M_{\odot}}$. If we increase the companion mass to be $0.4 \, {\rm M_{\odot}}$, we expect to measure up to $19$ MSP masses to this accuracy. Regarding this, the maximum angles that we expect the systems to be inclined at are $\sim 84^{\circ}$ and $79^{\circ}$ for the MSP systems with companion masses of $0.2 \, {\rm M_{\odot}}$ and $0.4 \, {\rm M_{\odot}}$, respectively. Therefore, the fraction of the MSPs in a MSP sample for which we expect to measure their masses to this accuracy using the Shapiro delay method is low.

\subsection{The mass distribution for WD companions to pulsars}
\begin{figure*}
\centering
 \includegraphics[width=0.45\textwidth]{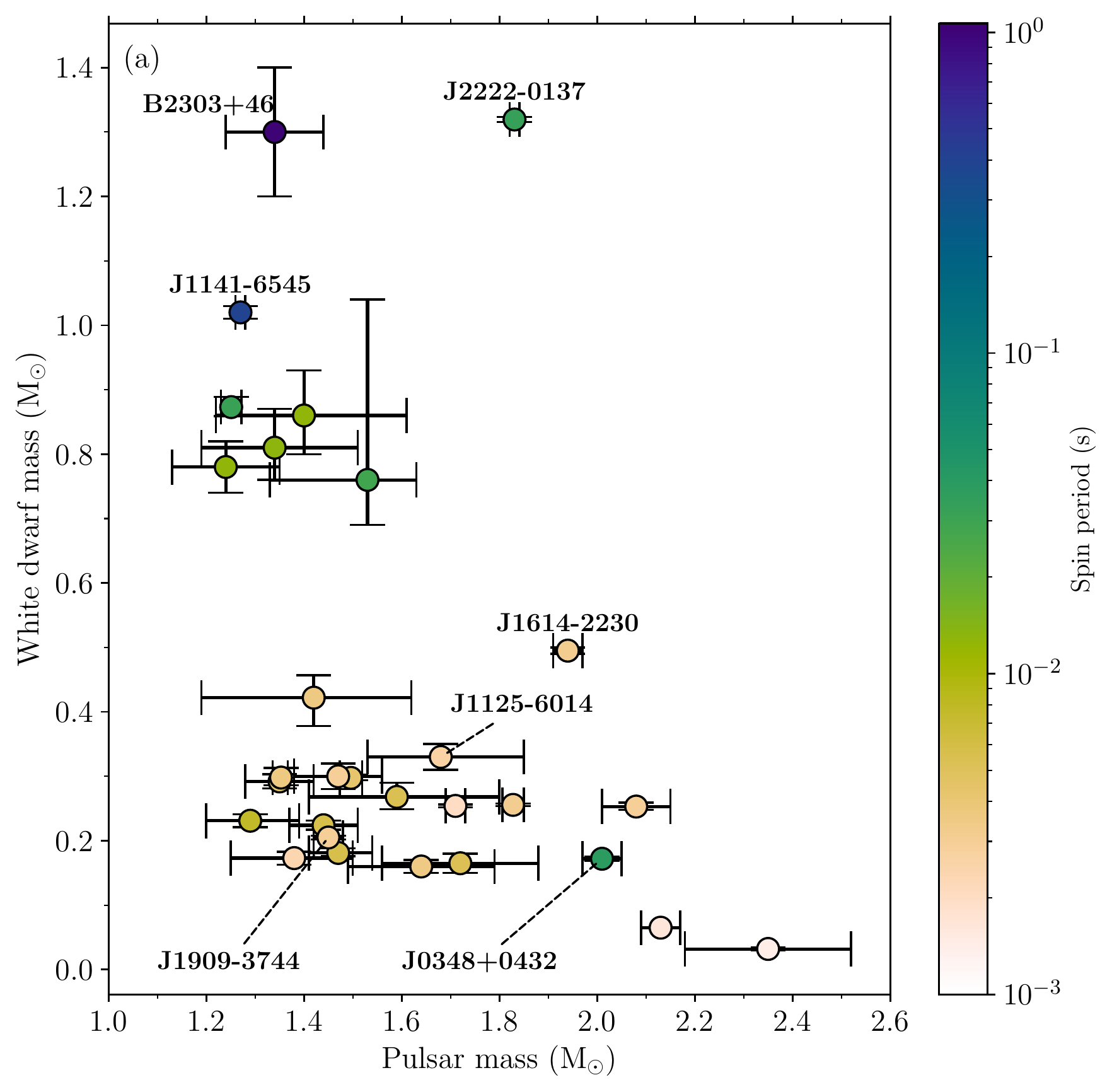}
 \includegraphics[width=0.45\textwidth]{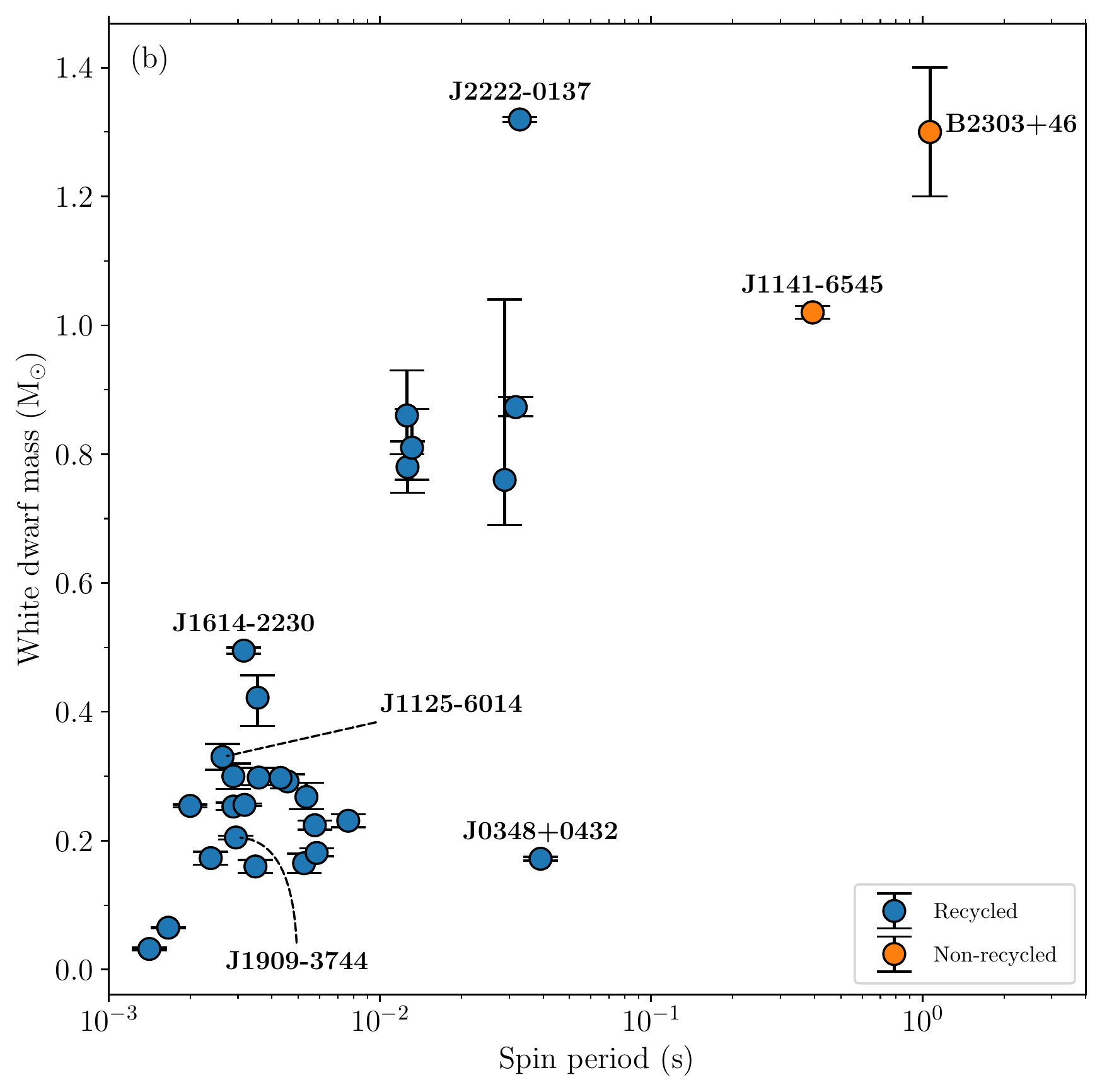}
\caption{Precise mass measurements for all known pulsar-WD systems located outside of globular clusters. (a): Pulsar masses vs. white dwarf (WD) companion masses with coloured circles indicating the spin period of pulsars. (b): WD companion masses versus spin periods of pulsars. The values of the pulsar and WD masses as well as their references are listed in Table \ref{tab:pulsar_wd_masses}, and the spin periods are taken from \texttt{psrcat}.}
\label{fig:pulsar_white_dwarf}
\end{figure*}

For all known pulsars that are located outside of globular clusters and whose companions are found to be WDs, we have plotted the precisely measured masses (with the pulsar mass uncertainty less than $15\%$), with logarithmic scaled coloured circles indicating the pulsar spin period, in the panel (a) of Figure \ref{fig:pulsar_white_dwarf}, as an update on Figure 6 in \cite{mfb+20}.

In that reference, they indicated two main populations in their plot: 1) systems with companions less massive than $0.46 \, {\rm M_{\odot}}$ as low-mass helium WD stars 2) systems with companion masses between $0.7$ - $0.9 \, {\rm M_{\odot}}$ as carbon oxygen WD stars. One of the conclusions is that there is a mass gap between $0.41$ and $0.7 \, {\rm M_{\odot}}$. In addition to these, there are five special systems with less common types of evolution.

First, in the case of PSR~J1141$-$6545 and B2303$+$46, the WDs formed before the pulsar; this is the reason why these systems have eccentric orbits as their pulsars have not been recycled \citep{vk99,ts00}, and are therefore significantly slower (see panel (b) of Figure \ref{fig:pulsar_white_dwarf}).
The remaining three systems have low-eccentricity systems and recycled pulsars. As discussed previously, PSR~J1614$-$2230 is a rare system with a heavy neutron star and a carbon-oxygen WD, thought to have been formed in Case A RLO \citep{tlk11}. PSR~J0348$+$0432 is a massive pulsar with a low-mass WD companion, has a tight orbit of $2.4$ hr, and the WD probably did not have an envelope mass less than the critical limit for hydrogen fusion, based on the assumption of \citet{afw+13}. This pulsar is unusually slow for such a tight orbit and such a low-mass companion. Finally,  PSR~J2222$-$0137 has a very massive ($M_{\rm WD} = 1.3194 \pm 0.0040\, \rm M_{\odot}$, \citealt{gfg+21}) and cool \citep{kbd+14} WD companion.

We have explored where PSRs~J1125$-$6014, J1614$-$2230, and J1909$-$3744 lie in Fig \ref{fig:pulsar_white_dwarf} relative to the other pulsars. They are very consistent with the established distribution of the pulsars and WDs as discussed in \cite{mfb+20}. PSR~J1125$-$6014 is starting to fill an empty region of parameter space close to PSR~J1614$-$2230, possibly because it also formed via Case A RLO.

Figure \ref{fig:pulsar_white_dwarf} highlights the apparent gap between the masses of the MSPs with $0.15$-$0.5 \, {\rm M_{\odot}}$ (this is even wider if we exclude binaries formed in Case A RLO), and those between $0.7$ and $0.9 \, {\rm M_{\odot}}$. Are there WD companions to neutron stars with masses in between?
If not, this would suggest that we have a higher probability of measuring a Shapiro delay when the mass function yields $0.7 < M_{\rm c} < 0.9 \, \rm M_{\odot}$, assuming $i = 90^\circ$. However, it is important to attempt to measure the Shapiro delay in systems where the mass function suggests intermediate masses in order to test the idea that there is indeed a gap in WD masses between $0.4$ and $0.7 \, \rm M_{\odot}$.

In panel (b) of Figure \ref{fig:pulsar_white_dwarf}, we have also plotted the precisely measured companion mass versus pulsar spin period. For the bulk of the recycled pulsars there is a strong correlation between
the companion mass and the log of the spin period. This is expected from basic considerations of stellar evolution: the more massive WD companions had more massive progenitors, which have evolved faster. Because of this faster evolution, the time available for mass transfer and accretion is much smaller; resulting in smaller ablation of the B-field and a smaller amount of spin-up. The exceptions are PSR~J0348$+$0242 and, as discussed before, PSR~J1614$-$2230.

Sometimes extending the timing baseline with archival data can lead to
increased accuracy of parameters that are undergoing secular changes, 
like position (via proper motion) and the time derivative of the
projected semi-major axis. These can help limit the inclination
angle from equation \ref{eq:xdot}.
We added $817$ ToAs from $8.34$ years of PPTA DR2e data sets of PSR~J1732$-$5049 \citep{rsc+21} with an rms residuals of $4.471 \, \mu \rm s$, to the current $4486$ MeerKAT ToAs with an rms residuals of $2.3 \, \mu$s in order to try and constrain $\dot{x}$ and find an upper limit on inclination angle; however, by using $18.97$ years of data we could only gain an upper limit on the measured $\dot{x} = 4(7) \times 10^{-16} $, so we were still not be able to achieve this.

\section{Conclusions}\label{sec:conclusions}

We have presented the timing analysis of seven MSPs using $\sim3$ years of MeerKAT observations, and provided  updated timing models. 
We detected evidence for the relativistic Shapiro delay in the timing residuals of all MSPs. By conducting a Bayesian analysis, we were able to measure the Shapiro delay using the orthometric amplitude $h_{3}$ and orthometric ratio $\varsigma$. For PSRs~J1125$-$6014, J1614$-$2230 and J1909$-$3744, we could constrain the mass of the pulsars to be $M_{\rm p} = 1.68^{+0.17}_{-0.15} \, {\rm M_{\odot}}$, $M_{\rm p} = 1.94 \pm 0.03 \, {\rm M_{\odot}}$ and $M_{\rm p} = 1.45 \pm 0.03 \, {\rm M_{\odot}}$, and the companion masses to be $M_{\rm c} = 0.33 \pm 0.02 \, {\rm M_{\odot}}$, $M_{\rm c} = 0.495 \pm 0.005 \, {\rm M_{\odot}}$ and $M_{\rm c} = 0.205 \pm 0.003 \, {\rm M_{\odot}}$, respectively. For pulsars with no meaningful mass constraints, $68\%$ likelihood of the companion masses and the orbital inclination angles were calculated using $\chi^2$ analysis, assuming fixed pulsar masses of $1.2 \, {\rm M_{\odot}}$ and $2.0 \, {\rm M_{\odot}}$.
We find that our mass measurements are in agreement with previous mass measurements for the same systems, and with the previously observed distribution of pulsar and WD masses.

We performed two series of simulations for exploring the change of errors in the Shapiro parameters, $\sigma_{\rm \sin i}$ and $\sigma_{\rm M_{\rm c}}$, at different orbital inclination angles, and derived the estimated errors in the pulsar mass $\sigma_{\rm M_{\rm p}}$. We then modeled the change of $\sigma_{\rm M_{\rm p}}$ as a function of the cosine of the orbital inclination angle. This model works well for $\cos i < 0.35$. In addition, using our simulations, we demonstrated that, at the likely inclination angles of the four systems, it was no surprise that their pulsar masses could not be determined. We provided an empirical polynomial formula that observers can use to estimate their likelihood of determining pulsar masses given good orbital coverage and the number and rms of timing residuals as a function of the orbital inclination angle.

\section*{Acknowledgements}
The MeerKAT telescope is operated by the South African Radio Astronomy Observatory, which is a facility of the National Research Foundation, an agency of the Department of Science and Innovation. 
MS, MB, DR, and RMS acknowledge support through the Australian Research Council Centre of Excellence for Gravitational Wave Discovery (OzGrav), through project number CE17010004.
RMS acknowledges support through Austarlian Research Council Future Fellowship FT190100155.
MK acknowledges significant support from the Max-Planck Society (MPG) and the MPIfR contribution to the PTUSE hardware.
MCB, DJC, PCCF, AP and VVK acknowledges continuing support from MPG.
AP acknowledges the using resources from the research grant ``iPeska'' (P.I. Andrea Possenti) funded under the INAF national call Prin-SKA/CTA approved with the Presidential Decree 70/2016. AP also had the support from the Ministero degli Affari Esteri e della Cooperazione Internazionale - Direzione Generale per la Promozione del Sistema Paese - Progetto di Grande Rilevanza ZA18GR02.
This work used the OzSTAR national facility at Swinburne University of Technology. OzSTAR is funded by Swinburne University of Technology and the National Collaborative Research Infrastructure Strategy (NCRIS).
This research has made use of NASA's Astrophysics Data System and software such as: \textsc{psrchive} \citep{vdo12}, \textsc{tempo2} \citep{hem06,ehm06}, \textsc{temponest} \citep{lah+14}, \textsc{pulseportraiture} \citep{pdr14,p19}, \textsc{numpy} \citep{numpy}, \textsc{scipy} \citep{scipy}, \textsc{matplotlib} \citep{matplotlib}, \textsc{ipython} \citep{ipython}, \textsc{astropy} \citep{astropy,astropy_v2}, \textsc{scikit-learn} \citep{scikit-learn}, \texttt{corner.py} \citep{corner16}, and  \textsc{cmasher} \citep{van20}.

\section*{Data Availability}
Data is available from the Swinburne pulsar portal: https://pulsars.org.au



\bibliographystyle{mnras}
\bibliography{main} 




\appendix

\section{pulsar ephemerides}\label{sec:ephemerides}

Tables of pulsar observational and derived parameters. The standard error on the last quoted digit is indicated by the values in parentheses.

\begin{table*}
\caption{Observed and derived timing model parameters for binary pulsars J0101$-$6422, J1101$-$6424, J1125$-$6014.}
\begin{tabular}{lllll}
\hline
Pulsar & \psra\ & \psrb\ & \psrg & \\
\hline
\multicolumn{5}{l}{General parameters}\\
\hline
Reference epoch (MJD) & $59130$ & $59132$ & $59174$ &\\

MJD range & $58595-59666$ & $58575-59689$ & $58589-59758$ &  \\

Fit $\chi^2$ / number of degrees of freedom &  $3018.65/3056$ & $7574.58/7612$ & $979.38/1027$ & \\

Post-fit RMS of residuals ($\mu$s) & $1.602$ & $7.973$ & $0.336$ & \\

L-band observation time ($h$) & $14.37$ & $39.17$ & $11.59$ & \\

UHF observation time ($h$) & $11.84$ & $-$ & $-$ & \\

Total number of observations L-band/UHF & $35/10$ & $80/0$ & $37/0$ & \\
\hline
\multicolumn{5}{l}{Spin and astrometric parameters}\\
\hline
Right ascension, $\alpha$ (epoch J2000; hh:mm:ss.s) & $01$:$01$:$11.13637(5)$ & $11$:$01$:$37.19156(4)$ & $11$:$25$:$55.243407(4)$ & \\

Declination, $\delta$ (epoch J2000; dd:mm:ss.s) & $-64$:$22$:$30.2768(4)$ & $-64$:$24$:$39.33311(17)$  & $-60$:$14$:$06.81026(3)$ & \\

Parallax, $\pi$ (mas) & $4.5(17)$ & $-$ & $1.6(2)$ & \\

Proper motion in RA, $\mu_{\rm \alpha} \cos \delta$ (mas yr$^{-1}$) &   $13.2(4)$ & $-1.3(2)$  &  $11.07(3)$ & \\

Proper motion in DEC, $\mu_{\rm \delta}$ (mas yr$^{-1}$) & $-9.6(4)$  & $-0.1(3)$  &  $-13.08(3)$ & \\

Total proper motion, $\mu$ (mas yr$^{-1}$) & $16.3(5)$ & $1.4(3)$ & $17.13(3)$ & \\

Spin period, $P$ (ms) & $2.57315201365217(9)$ & $5.10927298405781(12)$ & $2.630380782509540(5)$ & \\

First derivative of spin period, $\dot{P}$ ($10^{-20}$ s s$^{-1}$) & $0.5083(6)$ & $0.2591(7)$ & $0.37303(3)$ & \\

Dispersion measure, $DM$ (cm$^{-3}$ pc) & $11.9217(4)$ & $207.3647(5)$ & $52.93371(6)$ & \\

First derivative of DM, $\dot{DM}$ (cm$^{-3}$ pc s$^{-1}$) & $0.0005(5)$ & $-0.0009(3)$ & $-0.00105(3)$ & \\

Second derivative of DM, $\ddot{DM}$ (cm$^{-3}$ pc s$^{-2}$) & $-0.0010(11)$ & $0.0000(12)$ & $0.00186(15)$ & \\
\hline
\multicolumn{5}{l}{Orbital parameters}\\
\hline
Orbital period, $P_b$ (d)    & $1.7875967338(5)$ & $9.6117088579(9)$ & $8.75260365059(15)$ & \\

Projected semimajor axis of binary orbit, $x$ (lt-s)    & $1.7010463(2)$ & $14.02463520(16)$ & $8.33919893(3)$ & \\

Time of ascending node, $T_{\rm asc}$ (MJD)      & $59100.47696927(7)$ & $59081.94129419(3)$ & $59167.119295153(4)$ & \\

First Laplace-Lagrange parameter, $\epsilon_1$ ($10^{-6}$)   & $-0.3(2)$ & $21.57(3)$ & $0.512(6)$ & \\

Second Laplace-Lagrange parameter, $\epsilon_2$ ($10^{-6}$)  & $0.58(15)$ & $3.94(3)$ & $-0.612(5)$ & \\

Orthometric amplitude, $h_{3}$ ($\mu$s)       & $0.49(11)$ & $0.32(12)$ & $0.84(2)$ & \\

Orthometric ratio, $\varsigma$       & $0.66(8)$ & $0.81(15)$ & $0.804(11)$ & \\
\hline
\multicolumn{5}{l}{Stochastic parameters}\\
\hline
Log$_{10}$[EFAC] KAT receiver  & $0.008(10)$ & $-0.010(9)$ & $0.05(2)$ & \\

Log$_{10}$[EFAC] UHF receiver  & $0.030(9)$ & $-$  & $-$ & \\

Log$_{10}$[EQUAD] KAT receiver & $-8.3(1.0)$ & $-5.57(16)$ & $-6.69(4)$ & \\

Log$_{10}$[EQUAD] UHF receiver & $-7.9(1.2)$ & $-$ & $-$ &  \\

Log$_{10}$[ECORR] KAT receiver & $-8.5(9)$ & $-8.5(9)$ & $-8.4(8)$ &  \\

Log$_{10}$[ECORR] UHF receiver & $-8.8(8)$ & $-$ & $-$ & \\

Log$_{10}$[Red Amp] & $-$ & $-$ & $-$ &  \\

Red Index & $-$ & $-$ & $-$ &  \\

Log$_{10}$[DM Amp] & $-11.15(18)$ & $-10.94(13)$ & $-11.74(11)$ &  \\

DM Index & $2.1(7)$ & $2.8(7)$ & $1.6(5)$ &  \\
\hline
\end{tabular}
\label{tab:timing_solution_1}
\end{table*}

\begin{table*}
\caption{Observed and derived timing model parameters for binary pulsars J1514$-$4946, J1614$-$2230, J1732$-$5049, J1909$-$3744}
\begin{tabular}{lllll}
\hline
Pulsar & \psrc & \psrd & \psre & \psrf\\
\hline
\multicolumn{5}{l}{General parameters}\\
\hline
Reference epoch (MJD) & $59242$ & $59099$ & $59067$ & $59059$\\

MJD range & $58834-59651$ & $58589-59691$ & $58526-59608$ & $58526-59592$ \\

Fit $\chi^2$ / number of degrees of freedom & $461.05/463$ & $1673.19/1740$ & $4437.87/4467$ & $3523.40/2936$ \\

Post-fit RMS of residuals ($\mu$s) & $1.487$ & $0.648$ & $2.300$ & $0.090$ \\

L-band observation time ($h$) & $11.59$ & $7.02$ & $18.79$ & $23.66$ \\

UHF observation time ($h$) & $-$ & $-$ & $2.84$ & $-$ \\

Total number of observations L-band/UHF & $37/0$ & $69/0$ & $76/5$ & $187/0$ \\
\hline
\multicolumn{5}{l}{Spin and astrometric parameters}\\
\hline
Right ascension, $\alpha$ (epoch J2000; hh:mm:ss.s) & $15$:$14$:$19.10584(4)$ & $16$:$14:36.50968(4)$ & $17$:$32$:$47.766091(11)$ & $19$:$09$:$47.424651(5)$ \\

Declination, $\delta$ (epoch J2000; dd:mm:ss.s) & $-49$:$46$:$15.6363(4)$ & $-22$:$30$:$31.539(3)$ & $-50$:$49$:$00.3137(3)$ & $-37$:$44$:$14.91316(19)$ \\

Parallax, $\pi$ (mas) & $-$ & $1.11(17)$ & $0.8(6)$ & $0.75(8)$ \\

Proper motion in RA, $\mu_{\rm \alpha} \cos \delta$ (mas yr$^{-1}$) & $-7.8(2)$ & $4.9(7)$ & $-0.62(17)$ & $-9.60(5)$ \\

Proper motion in DEC, $\mu_{\rm \delta}$ (mas yr$^{-1}$) & $-10.3(4)$ & $-26(4)$ & $-9.6(4)$ & $-35.6(2)$ \\

Total proper motion, (mas yr$^{-1}$) & $13.0(4)$ & $27(4)$ & $9.6(5)$ & $36.8(3)$ \\

Spin period, $P$ (ms) & $3.58933674132885(6)$ & $3.15080765799915(4)$ & $5.31255029615166(7)$ & $2.947108074685310(15)$ \\

First derivative of spin period, $\dot{P}$ ($10^{-20}$ s s$^{-1}$) & $1.8670(10)$ & $0.9625(4)$ & $1.4182(5)$ & $1.4023(3)$ \\

Dispersion measure, $DM$ (cm$^{-3}$ pc) & $31.00853(13)$ & $34.48755(14)$ & $56.82189(13)$ & $10.39090(5)$ \\

First derivative of DM, $\dot{DM}$ (cm$^{-3}$ pc s$^{-1}$) & $-0.00016(15)$ & $-0.00054(8)$ & $-0.00146(16)$ & $0.000053(12)$ \\

Second derivative of DM, $\ddot{DM}$ (cm$^{-3}$ pc s$^{-2}$) & $0.0007(5)$ & $-0.0003(3)$ & $0.0009(4)$ & $-0.00020(14)$ \\
\hline
\multicolumn{5}{l}{Orbital parameters}\\
\hline
Orbital period, $P_b$ (d)    & $1.9226535542(3)$ & $8.6866194264(4)$ & $5.2629972188(15)$ & $1.533449477320(12)$ \\

Projected semimajor axis of binary orbit, $x$ (lt-s)    & $1.9332622(3)$ & $11.29120702(7)$ & $3.98286955(19)$ & $1.897994789(14)$ \\

Time of ascending node, $T_{\rm asc}$ (MJD)      & $59211.98535052(7)$ & $59098.046639722(11)$ & $59064.55306932(4)$ & $59057.600910299(3)$ \\

First Laplace-Lagrange parameter, $\epsilon_1$ ($10^{-6}$)   & $4.0(3)$ & $0.930(12)$ & $1.65(9)$ & $1.940(14)$ \\

Second Laplace-Lagrange parameter, $\epsilon_2$ ($10^{-6}$)  & $3.1(3)$ & $-1.297(17)$ & $-8.02(6)$ & $-0.104(14)$ \\

Orthometric amplitude, $h_{3}$ ($\mu$s)       & $0.5(2)$ & $2.340(19)$ & $0.53(13)$ & $0.845(7)$ \\

Orthometric ratio, $\varsigma$       & $0.81(13)$ & $0.9858(3)$ & $0.4(3)$ & $0.9441(16)$ \\
\hline
\multicolumn{5}{l}{Stochastic parameters}\\
\hline
Log$_{10}$[EFAC] KAT receiver & $-0.044(15)$ & $-0.003(8)$ & $0.008(8)$ & $0.010(7)$\\

Log$_{10}$[EFAC] UHF receiver  & $-$ & $-$ & $-0.070(14)$ & $-$\\

Log$_{10}$[EQUAD] KAT receiver & $-8.3(1.0)$ & $-8.6(9)$ & $-6.00(3)$ & $-7.75(8)$\\

Log$_{10}$[EQUAD] UHF receiver & $-$ & $-$ & $-8.2(1.1)$ & $-$\\

Log$_{10}$[ECORR] KAT receiver & $-8.7(8)$ & $-7.7(5)$ & $-8.9(7)$ & $-7.57(4)$\\

Log$_{10}$[ECORR] UHF receiver & $-$ & $-$ & $-8.6(9)$ & $-$\\

Log$_{10}$[Red Amp] & $-$ & $-13.3(4)$ & $-$ & $-13.19(15)$\\

Red Index & $-$ & $3.9(1.3)$ & $-$ & $2.6(6)$\\

Log$_{10}$[DM Amp] & $-$ & $-11.41(11)$ & $-11.39(9)$ & $-11.83(9)$\\

DM Index & $-$ & $1.4(4)$ & $0.16(14)$ & $1.8(3)$\\
\hline
\end{tabular}
\label{tab:timing_solution_2}
\end{table*}

\section{Pulsar and companion mass measurements with the relevant references}

\begin{table*}
\caption{Pulsars with estimated WD companion masses plotted in Fig \ref{fig:pulsar_white_dwarf}}
\begin{tabular}{llll}
\hline
Pulsar & $M_{\rm p}$ & $M_{\rm c}$ & Reference \\
\hline
J0348$+$0432 & $2.01^{0.04}_{0.04}$ & $0.172^{0.003}_{0.003}$ & \cite{afw+13} \\
J0437$-$4715 & $1.44^{0.07}_{0.07}$ & $0.224^{0.007}_{0.007}$ & \cite{rhc+16} \\
J0621$+$1002 & $1.53^{0.1}_{0.2}$ & $0.76^{0.28}_{0.07}$ & \cite{kas12} \\
J0740$+$6620 & $2.08^{0.07}_{0.07}$ & $0.253^{0.006}_{0.005}$ & \cite{fcp+21} \\
J0751$+$1807 & $1.64^{0.15}_{0.15}$ & $0.16^{0.01}_{0.01}$ & \cite{dcl+16} \\
J0952$-$0607 & $2.35^{0.17}_{0.17}$ & $0.032^{0.002}_{0.002}$ & \cite{rkf+22} \\
J0955$-$6150 & $1.71^{0.02}_{0.02}$ & $0.254^{0.002}_{0.002}$ & \cite{svf+22} \\
J1012$+$5307 & $1.72^{0.16}_{0.16}$ & $0.165^{0.015}_{0.015}$ & \cite{miv+20} \\
J1125$-$6014 & $1.68^{0.17}_{0.15}$ & $0.33^{0.02}_{0.02}$ & This work \\
J1141$-$6545 & $1.27^{0.01}_{0.01}$ & $1.02^{0.01}_{0.01}$ & \cite{bbv08} \\
J1614$-$2230 & $1.94^{0.03}_{0.03}$ & $0.495^{0.005}_{0.005}$ & This work \\
J1713$+$0747 & $1.35^{0.07}_{0.07}$ & $0.292^{0.011}_{0.011}$ & \cite{abb+18} \\
J1738$+$0333 & $1.47^{0.07}_{0.06}$ & $0.181^{0.007}_{0.005}$ & \cite{avk+12} \\
J1802$-$2124 & $1.24^{0.11}_{0.11}$ & $0.78^{0.04}_{0.04}$ & \cite{fsk+10} \\
J1810$+$1744 & $2.13^{0.04}_{0.04}$ & $0.065^{0.001}_{0.001}$ & \cite{rkf+21} \\
B1855$+$09 & $1.37^{0.13}_{0.10}$ & $0.244^{0.014}_{0.012}$ & \cite{abb+18} \\
J1909$-$3744 & $1.45^{0.03}_{0.03}$ & $0.205^{0.003}_{0.003}$ & This work \\
J1918$-$0642 & $1.29^{0.1}_{0.09}$ & $0.231^{0.010}_{0.010}$ & \cite{abb+18} \\
J1933$-$6211 & $1.42^{0.20}_{0.23}$ & $0.422^{0.035}_{0.044}$ & Geyer et al. 2022 (submitted) \\
J1946$+$3417 & $1.828^{0.022}_{0.022}$ & $0.2556^{0.0019}_{0.0019}$ & \cite{bfk+17} \\
J1949$+$3106 & $1.34^{0.17}_{0.15}$ & $0.81^{0.06}_{0.05}$ & \cite{zfk+19} \\
J1950$+$2414 & $1.496^{0.023}_{0.023}$ & $0.2975^{0.0046}_{0.0038}$ & \cite{zfk+19} \\
J2043$+$1711 & $1.38^{0.12}_{0.13}$ & $0.173^{0.010}_{0.010}$ & \cite{abb+18} \\
J2045$+$3633 & $1.251^{0.021}_{0.021}$ & $0.873^{0.016}_{0.014}$ & \cite{mfb+20} \\
J2053$+$4650 & $1.40^{0.21}_{0.18}$ & $0.86^{0.07}_{0.06}$ & \cite{bcf+17} \\
J2222$-$0137 & $1.831^{0.010}_{0.010}$ & $1.3194^{0.0040}_{0.0040}$ & \cite{gfg+21} \\
J2234$+$0611 & $1.353^{0.014}_{0.017}$ & $0.298^{0.015}_{0.012}$ & \cite{sfa+19} \\
B2303$+$46 & $1.34^{0.10}_{0.10}$ & $1.30^{0.10}_{0.10}$ & \cite{tc99} \\
J2339$-$0533 & $1.47^{0.09}_{0.09}$ & $0.30^{0.02}_{0.02}$ & \cite{krf+20} \\
\hline
\end{tabular}
\label{tab:pulsar_wd_masses}
\end{table*}

\section{}\label{sec:probability_distributions}
Posterior probability distributions of parameters. The figures are created using \texttt{corner.py} developed by \citet{corner16}.

\begin{figure*}
\centering
 \includegraphics[width=\textwidth]{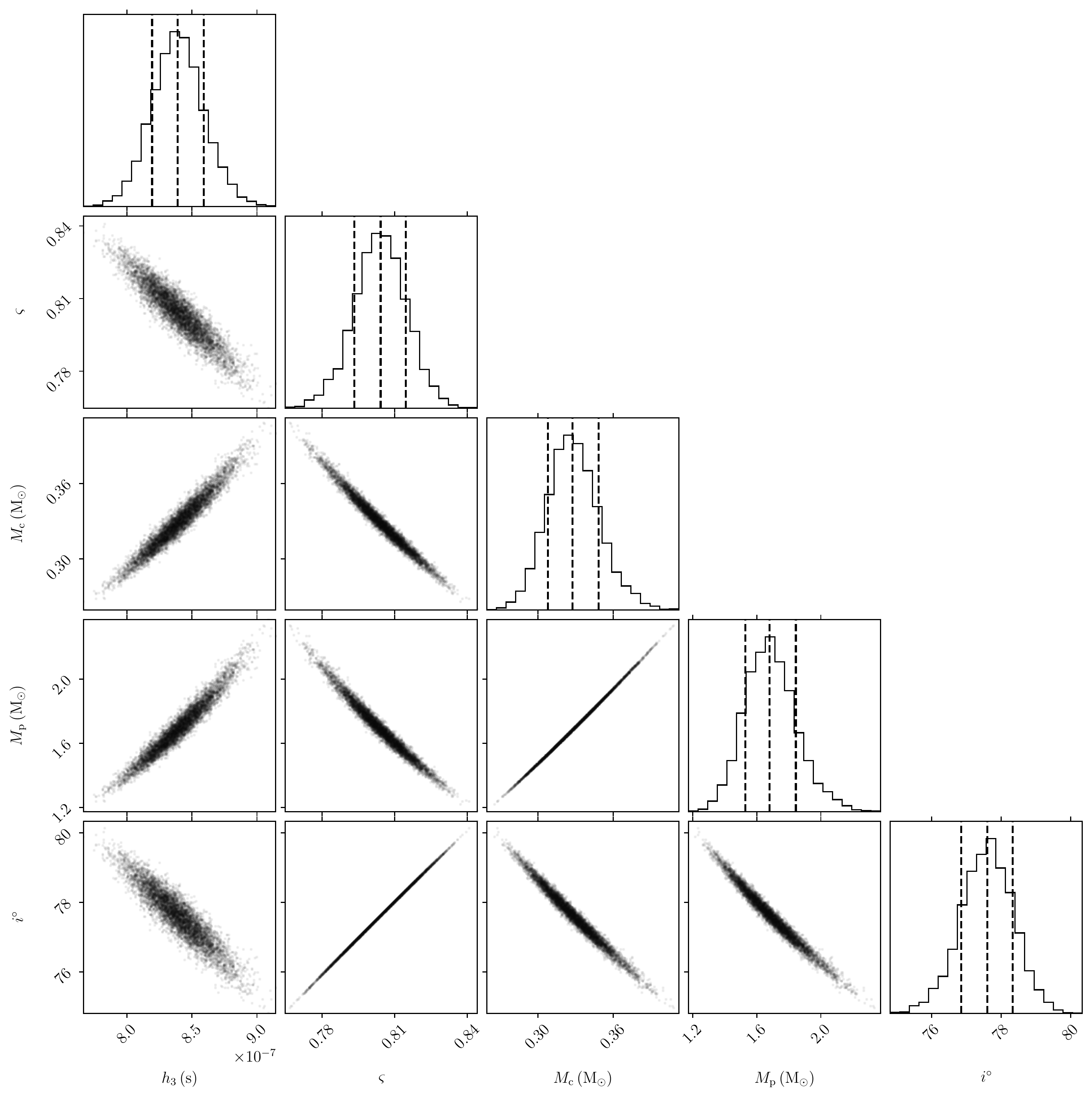}
\caption{2D marginalized posterior probability distributions of the timing parameters: $h_{3}$ and $\varsigma$, and the derived parameters: $M_{\rm c}$, $M_{\rm p}$, $i$, for J1125$-$6014. The dashed vertical lines on the 1-D histograms are showing the median and $68\%$ confidence levels.}
\label{fig:J1125_PDF}
\end{figure*}

\begin{figure*}
\centering
 \includegraphics[width=\textwidth]{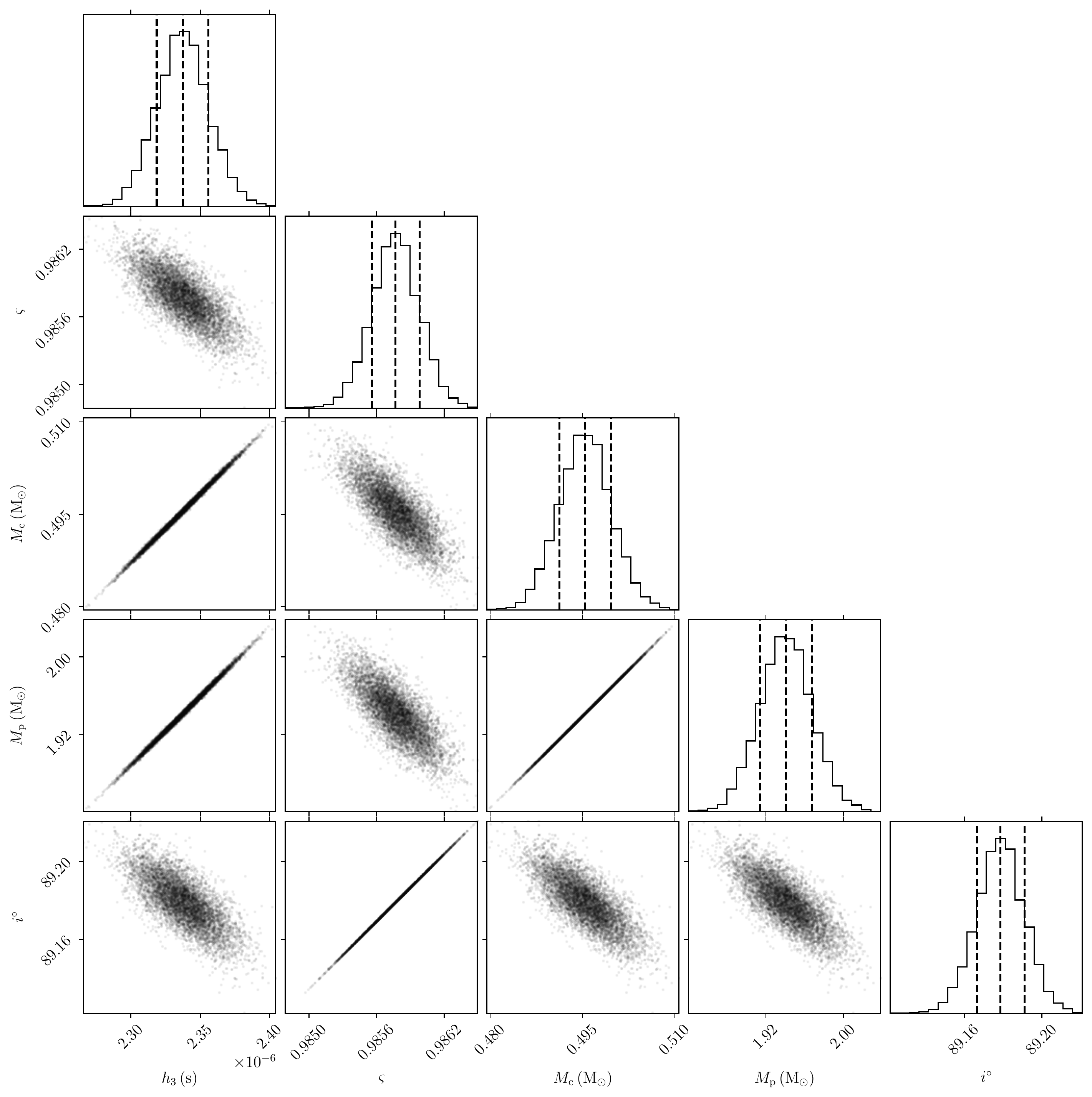}
\caption{2D marginalized posterior probability distributions of the timing parameters: $h_{3}$ and $\varsigma$, and the derived parameters: $M_{\rm c}$, $M_{\rm p}$, $i$, for J1614$-$2230. The dashed vertical on the 1-D histograms lines are showing the median and $68\%$ confidence levels..}
\label{fig:J1614_PDF}
\end{figure*}

\begin{figure*}
\centering
 \includegraphics[width=\textwidth]{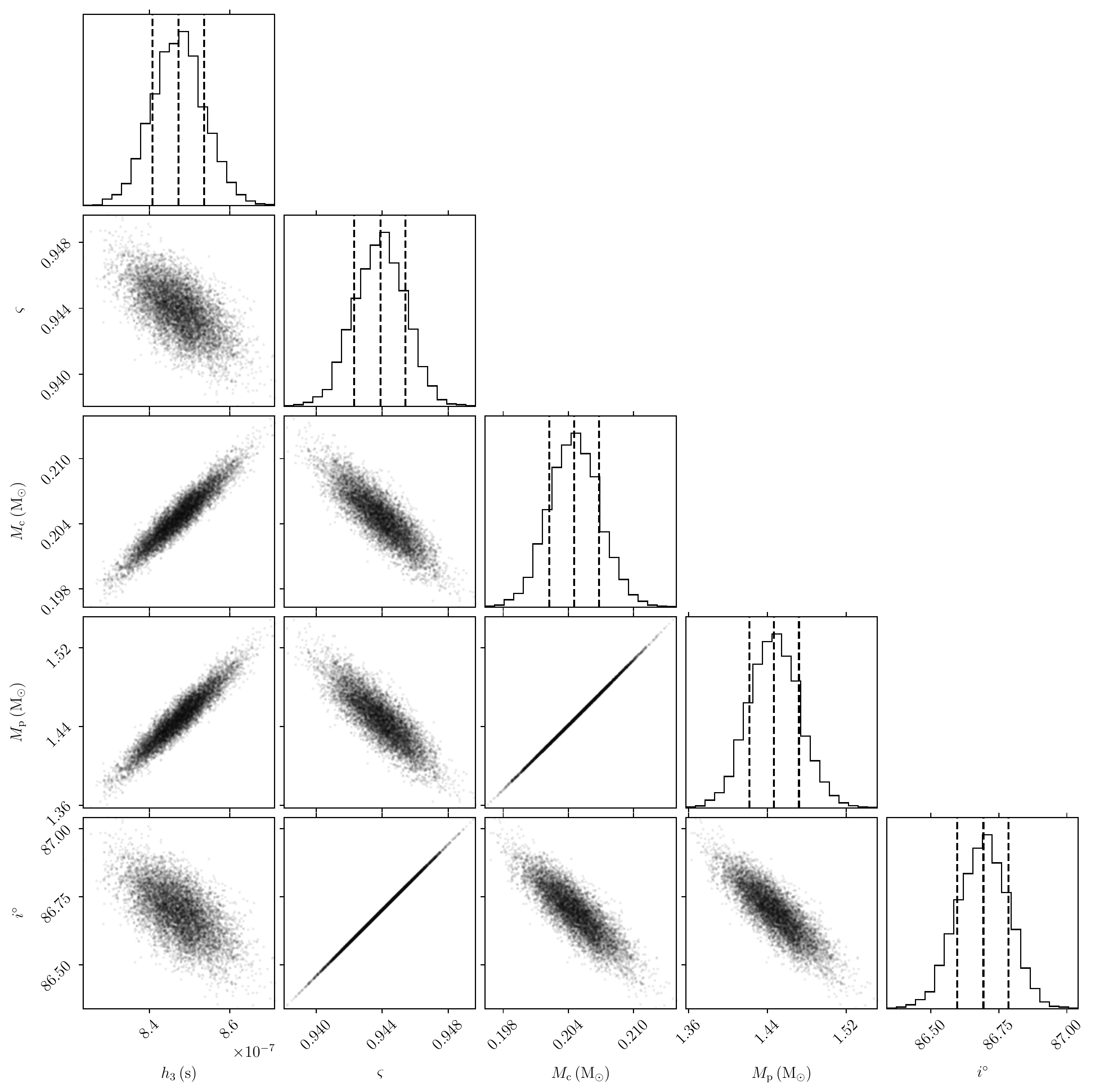}
\caption{2D marginalized posterior probability distributions of the timing parameters: $h_{3}$ and $\varsigma$, and the derived parameters: $M_{\rm c}$, $M_{\rm p}$, $i$, for J1909$-$3744. The dashed vertical lines on the 1-D histograms are showing the median and $68\%$ confidence levels.}
\label{fig:J1909_PDF}
\end{figure*}


\bsp	
\label{lastpage}
\end{document}